\documentclass[published]{JHEP}
\JHEP{06(1999)017}
\usepackage{epsfig}

\title{$B\to X_u\,l\,\bar\nu_l$ decay distributions to 
order $\alpha_s$}

\author{Fulvia De Fazio\\
	Istituto Nazionale di Fisica Nucleare -- Sezione di Bari\\
	via Amendola n.~173, 70126 Bari, Italy\\
	E-mail: \email{Fulvia.Defazio@ba.infn.it}}

\author{Matthias Neubert\\
	Stanford Linear Accelerator Center, Stanford University\\
	Stanford, California 94309, U.S.A.\\
	E-mail: \email{neubert@slac.stanford.edu}}

\abstract{An analytic result for the $O(\alpha_s)$ corrections to the
triple differential $B\to X_u\,l\,\bar\nu_l$ decay rate is presented,
to leading order in the heavy-quark expansion. This is relevant for
computing partially integrated decay distributions with arbitrary cuts
on kinematic variables. Several double and single differential
distributions are derived, most of which generalize known results. In
particular, an analytic result for the $O(\alpha_s)$ corrections to
the hadronic invariant mass spectrum is presented. The effects of
Fermi motion, which are important for the description of decay spectra
close to infrared sensitive regions, are included. The behaviour of
perturbation theory in the region of time-like momenta is also
investigated.}

\received{May 15, 1999}
\accepted{June 15, 1999}

\begin{document}

\section{Introduction}

Inclusive semileptonic decays of $B$ mesons are of prime importance 
to determining the parameters $|V_{cb}|$ and $|V_{ub}|$ of the 
Cabibbo--Kobayashi--Maskawa matrix~\cite{BaBar}. They also 
serve as probes for physics beyond the Standard Model, such as an 
extended Higgs sector~\cite{Zoltan} or right-handed weak couplings of 
the $b$ quark~\cite{Volo,Tom}. The total decay rates for these 
processes can be calculated in a systematic expansion in inverse 
powers of $m_b$~\cite{Chay,Bigi,MaWe,Blok,Falk,Thom,review}. The same 
formalism can also be applied to calculate differential decay 
distributions, provided a sufficient sampling of hadronic final 
states is ensured by kinematics. Close to the boundary of phase 
space, the heavy-quark expansion must be generalized into a twist 
expansion to account for the effects of the ``Fermi motion'' of the 
$b$ quark inside the $B$ meson~\cite{me,Fermi}.

The leading contribution in the $1/m_b$ expansion is given by the 
free $b$-quark decay into partons, calculated in perturbation theory 
as a power series in $\alpha_s$. Several authors have computed 
radiative corrections to various semileptonic decay rates and
spectra. In particular, the QCD corrections to the total inclusive 
$B\to X_u\,l\,\bar\nu_l$ decay rate have recently been computed to
$O(\alpha_s^2)$~\cite{Rit}, and the rate for $B\to X_c\,l\,\bar\nu_l$ 
is known to the same order from an extrapolation of exact results
obtained at three different values of the invariant mass of the 
lepton--neutrino pair~\cite{Czar}. However, to our knowledge no 
results for the $O(\alpha_s)$ corrections to the fully differential 
$B\to X\,l\,\bar\nu_l$ decay distribution have been published so 
far. (An early investigation of these corrections was performed 
in~\cite{Martin}, where results are presented in complicated equations 
involving one-dimensional integrals.) Whereas this distribution by 
itself is not of direct phenomenological relevance (because it does 
not contain sufficient averaging over hadronic final states to be 
realistic), it is a necessary ingredient in the derivation of 
predictions for inclusive spectra with arbitrary cuts on kinematic 
variables. 

Here we present analytic results for the fully differential decay rate
and several double and single differential distributions for $B\to
X_u\,l\,\bar\nu_l$ decays. Throughout, we work to leading order in the
heavy-quark expansion, omitting corrections of order
$(\Lambda/m_b)^2$, which have been discussed by previous authors (see,
e.g.,~\cite{Bigi,MaWe,Blok,Falk,Thom,review}). The QCD corrections are
calculated including terms that do not contribute in the limit of
vanishing lepton mass, so that our results allow treating the case of
decays into $\tau$ leptons. Semileptonic decays into final states
containing a charm quark will be discussed elsewhere. A technical
complication arises from the presence of soft and collinear
singularities, which do not cancel in the fully differential decay
distribution. These unphysical singularities appear because the
calculation of inclusive rates is performed using external states
containing free quarks and gluons. By virtue of global quark--hadron
duality, the inclusive partonic rates are dual to the corresponding
hadronic rates if a sufficient averaging over many final states is
performed. We set $m_u=0$ and regulate the infrared singularities by
introducing a fictitious gluon mass $\lambda$. The limit $\lambda\to
0$ can be taken at the end of the calculation and leads to singular
distributions at $p^2=0$, where $p$ is the total parton
momentum. Inclusive spectra obtained by integration over a range in
$p^2$ are infrared finite.

\section{Hadronic tensor at next-to-leading order in $\alpha_s$}

All strong-interaction dynamics relevant to inclusive semileptonic
decays is encoded in the hadronic tensor
\begin{equation}
   W_{\mu\nu}(p,v) = -\frac{1}{\pi}\,\mbox{Im}\,T_{\mu\nu}(p,v) \,,
\end{equation}
where 
\begin{equation}\label{Tmunu}
   T_{\mu\nu}(p,v) = -i\int\mbox{d}^4x\,e^{i(p-m_b v)\cdot x}\,
   \frac{\langle B(v)|\,\mbox{T}\{J_\mu^\dagger(x),J_\nu(0)\}\,
         |B(v)\rangle}{2 M_B}
\end{equation}
is the forward scattering amplitude given by the $B$-meson matrix
element of the time-ordered product of two weak currents, with
$J_\mu=\bar u\gamma_\mu(1-\gamma_5)b$. For the calculation of QCD
corrections it is convenient to choose the $b$-quark velocity $v$
(which can be taken to coincide with the velocity of the $B$ meson)
and the total parton momentum $p$ as the two independent variables
characterizing the hadronic tensor. The total momentum carried by the
leptons is $q=m_b v-p$.

Because $v^2=1$, the two independent kinematic invariants are $v\cdot
p$ and $p^2$. The most general Lorentz-invariant decomposition of the
hadronic tensor contains five invariant functions $W_i(v\cdot p,p^2)$,
which we define as
\begin{eqnarray}
   W_{\mu\nu}(p,v) &=& W_1(v\cdot p,p^2) \Big( p_\mu v_\nu
    + p_\nu v_\mu - g_{\mu\nu}\,v\cdot p
    - i\epsilon_{\mu\nu\alpha\beta}\,p^\alpha v^\beta \Big)-
    \nonumber\\
   &&{}- W_2(v\cdot p,p^2)\,g_{\mu\nu}
    + W_3(v\cdot p,p^2)\,v_\mu v_\nu + \nonumber\\
   &&{}+ W_4(v\cdot p,p^2)\,(p_\mu v_\nu + p_\nu v_\mu) 
    + W_5(v\cdot p,p^2)\,p_\mu\,p_\nu \,.
\end{eqnarray}
At tree level $W_1=2\delta(p^2)$ and $W_{i\ne 1}=0$. The five
invariant functions $W_i$ suffice to calculate arbitrary semileptonic
decay distributions, including the case where the mass of the charged
lepton is not neglected. In general, these distributions can be
written in terms of three independent kinematic variables. One common
choice of such variables is the charged-lepton energy $E_l$, the total
lepton energy $v\cdot q$ (both defined in the $B$-meson rest frame),
and the invariant mass $q^2$ of the lepton pair. Here we choose a
different set of variables, because the hadronic tensor is most
conveniently calculated in terms of $v\cdot p$ and $p^2$. Besides,
experimentally the neutrino cannot be detected, whereas the total
invariant mass and energy of the hadronic final state can be
reconstructed directly. In terms of the parton variables, these
quantities are given by
\begin{equation}\label{rela}
   s_H = p^2 + 2\bar\Lambda\,v\cdot p+\bar\Lambda^2 \,,\qquad
   E_H = v\cdot p + \bar\Lambda \,,
\end{equation}
where $\bar\Lambda=M_B-m_b$. Experimental cuts on the region of low
hadronic invariant mass or energy have been suggested as efficient
ways to discriminate the small $B\to X_u\,l\,\bar\nu_l$ signal against
the much larger background from $B\to X_c\,l\,\bar\nu_l$
decays~\cite{Barg,Bouz,Greub,Dike,FLW}. Such a discrimination is
important for a reliable determination of the
Cabibbo--Kobayashi--Maskawa matrix element $|V_{ub}|$.

Although our results allow to treat the more general case, for the 
purpose of our phenomenological discussion we will set $m_l=0$. We 
introduce the scaling variables
\begin{equation}
   x = \frac{2 E_l}{m_b} \,,\qquad
   \hat p^2 = \frac{p^2}{m_b^2} \,,\qquad
   z = \frac{2 v\cdot p}{m_b} \,,
\end{equation}
in terms of which the triple differential decay rate is
\begin{eqnarray}\label{rate}
   \frac{\mbox{d}^3\Gamma}{\mbox{d}x\,\mbox{d}z\,\mbox{d}\hat p^2}
   &=& 12\Gamma_0\,\bigg\{ (1+\bar x-z)(z-\bar x-\hat p^2)\,
    \frac{m_b^2}{2}\,W_1 + (1-z+\hat p^2)\,\frac{m_b}{2}\,W_2 +
    \nonumber\\
   &&\hspace{1.1truecm}\mbox{}+ [\bar x(z-\bar x)-\hat p^2]\,
    \frac{m_b}{4}(W_3 + 2m_b W_4 + m_b^2 W_5) \bigg\} \,,
\end{eqnarray}
where $\bar x=1-x$, and
\begin{equation}
   \Gamma_0 = \frac{G_F^2 |V_{ub}|^2 m_b^5}{192\pi^3} \,.
\end{equation}
The phase space for these variables is
\begin{equation}\label{ps3}
   0\le x\le 1 \,,\qquad
   \bar x\le z\le 1+\bar x \,,\qquad
   \hbox{max}(0,z-1)\le \hat p^2\le \bar x(z-\bar x) \,.
\end{equation}
For fixed values of $z$ and $\hat p^2$ the lepton energy can vary in
the range $\frac 12(z-\sqrt{z^2-4\hat p^2})\le\bar x\le 
\frac 12(z+\sqrt{z^2-4\hat p^2})$, and since the hadronic tensor is 
independent of $x$ it is possible to integrate over this variable
to obtain the double differential spectrum
\begin{eqnarray}\label{double}
   \frac{1}{\Gamma_0}\,
   \frac{\mbox{d}^2\Gamma}{\mbox{d}z\,\mbox{d}\hat p^2}
   &=& 2\sqrt{z^2-4\hat p^2}\,\bigg\{ [z(3-2z)-\hat p^2(4-3z)]\,
    \frac{m_b^2}{2}\,W_1 + \nonumber\\
   &&\hspace{2.3truecm}\mbox{}+ 6(1-z+\hat p^2)\,
    \frac{m_b}{2}\,W_2 + \nonumber\\
   &&\hspace{2.3truecm}\mbox{}+ (z^2-4\hat p^2)\,\frac{m_b}{4}
    (W_3 + 2m_b W_4 + m_b^2 W_5) \bigg\} \,,
\end{eqnarray}
where 
\begin{equation}\label{ps}
   0\le z\le 2 \,,\qquad
   \hbox{max}(0,z-1)\le \hat p^2\le \frac{z^2}{4} \,.
\end{equation}

\EPSFIGURE{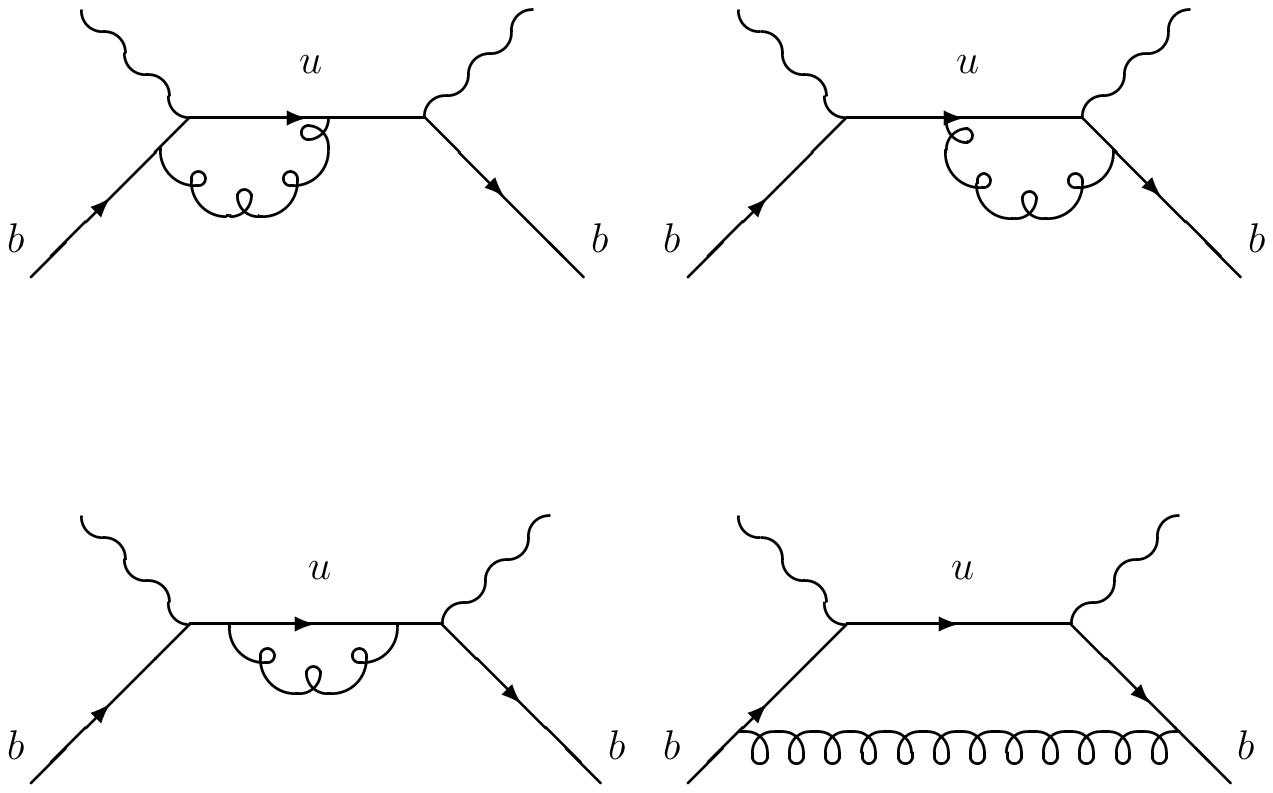,width=21em} {\label{fig:1}One-loop diagrams
contributing to the forward amplitude $T_{\mu\nu}(p,v)$.}

It is worth stressing at this point that any perturbative description 
of inclusive decay rates must necessarily be formulated in terms of 
scheme-dependent para\-meters such as the $b$-quark mass $m_b$ and the 
related parameter $\bar\Lambda$. These parameters are well-defi\-ned to 
a given order in perturbation theory. In the context of our 
perturbative analysis, $m_b$ is to be identified with the one-loop 
pole mass of the $b$ quark. The concept of a pole mass becomes 
ambiguous, however, if one goes beyond perturbation theory, and in 
that sense neither $m_b$ nor $\bar\Lambda$ are physical quantities. 
We will see in section~\ref{sec:fermi} that, for normalized inclusive 
decay spectra, all reference to these parameters disappears once the 
leading nonperturbative corrections are included in the heavy-quark 
expansion. 

The $O(\alpha_s)$ corrections to the hadronic tensor are obtained
by evaluating the contributions of all physical cuts of the diagrams
shown in figure~\ref{fig:1}, supplemented by wave-function 
renormalization graphs for the external $b$ quarks. The sum of 
all contributions is gauge independent and free of ultraviolet 
divergences. There are, however, infrared divergences for $p^2\to 0$ 
from soft and collinear gluons, which we regularize with a fictitious 
gluon mass $\lambda\equiv m_b\hat\lambda$. The limit $\hat\lambda\to 
0$ is taken whenever possible. We expand the invariant functions in a 
perturbation series as (with $C_F=4/3$)
\begin{equation}
   W_i(z,\hat p^2) = W_i^{(0)}(z,\hat p^2)
   + \frac{C_F\alpha_s}{4\pi}\,W_i^{(1)}(z,\hat p^2)
   + O(\alpha_s^2) \,,
\end{equation}
where $\frac{m_b^2}{2}\,W_1^{(0)}=\delta(\hat p^2)$, and all other 
functions vanish at leading order. Our results for the next-to-leading 
coefficients can be presented as follows:
\begin{eqnarray}\label{Wi}
   \frac{m_b^2}{2}\,W_1^{(1)} &=& - \delta(\hat p^2) \left[
    4\ln^2\!z - 6\ln z + \frac{2\ln z}{1-z} + 4 L_2(1-z) + \pi^2
    + \frac{15}{2} \right] - \nonumber\\
   &&{}- f_{\rm IR}(\hat\lambda^2,z,\hat p^2) + \frac{4}{\hat p^2} 
    \left[ \frac{1}{t} \ln\frac{1+t}{1-t} + \ln\frac{\hat p^2}{z^2}
    \right] + \nonumber\\
   &&{}+ 1 - \frac{(8-z)(2-z)}{z^2 t^2} + \left[
    \frac{2-z}{2z} + \frac{(8-z)(2-z)}{2z^2 t^2}
    \right] \frac{1}{t} \ln\frac{1+t}{1-t} \,, \nonumber\\
   \nonumber\\
   \frac{m_b}{2}\,W_2^{(1)} &=& \frac{8-z}{4}
    + \frac{32-8z+z^2}{4z t^2} - \left[ \frac{z t^2}{8} + \frac{4-z}{4}
    + \frac{32-8z+z^2}{8z t^2} \right] \frac{1}{t} \ln\frac{1+t}{1-t}
    \,, \nonumber\\
   \nonumber\\
   \frac{m_b}{4}\,W_3^{(1)} &=& - \frac{8-3z}{8}
    + \frac{32+22z-3z^2}{4z t^2} - \frac{3(12-z)}{8 t^4} + \nonumber\\
   &&{}+ \left[ \frac{z t^2}{16}
    + \frac{5(4-z)}{16} - \frac{64+56z-7z^2}{16z t^2}
    + \frac{3(12-z)}{16 t^4} \right] \frac{1}{t} \ln\frac{1+t}{1-t}
    \,, \nonumber\\
   \nonumber\\
   \frac{m_b^2}{2}\,W_4^{(1)} &=& \frac{2}{1-z} \left[
    \frac{z\ln z}{1-z} + 1 \right] \delta(\hat p^2) 
    - 1 - \frac{32-5z}{2z t^2} + \frac{3(12-z)}{2z t^4} - \nonumber\\
   &&{}- \left[ \frac{8-3z}{4z} - \frac{22-3z}{2z t^2} 
    + \frac{3(12-z)}{4z t^4} \right] \frac{1}{t}
    \ln\frac{1+t}{1-t} \,, \nonumber\\
   \nonumber\\
   \frac{m_b^3}{4}\,W_5^{(1)}
   &=& \frac{2}{1-z} \left[ \frac{1-2z}{1-z} \ln z - 1 \right]
    \delta(\hat p^2)  - \frac{8+z}{2z^2 t^2}
    - \frac{3(12-z)}{2z^2 t^4} + \nonumber\\
   &&{}+ \left[ \frac{1}{4z} - \frac{2-z}{2z^2 t^2} 
    + \frac{3(12-z)}{4z^2 t^4} \right] \frac{1}{t} 
    \ln\frac{1+t}{1-t} \,.
\end{eqnarray}
Here $t=\sqrt{1-4\hat p^2/z^2}$, $L_2(x)$ is the dilogarithm, and we 
have omitted the step function $\theta(\hat p^2)$ multiplying all 
regular terms. The expressions given above are regular for $t\to 0$, 
corresponding to the kinematic limit where $\hat p^2\to{z^2}/{4}$, 
i.e.~$|\vec p\,|=0$. 

The function $f_{\rm IR}(\hat\lambda^2,z,\hat p^2)$ entering the
expression for $W_1^{(1)}$ contains all reference to the infrared
regulator. It is given by
\begin{eqnarray}
   f_{\rm IR}(\hat\lambda^2,z,\hat p^2)
   &=& \delta(\hat p^2) \left[ \ln^2\!\hat\lambda^2
    + (5-4\ln z)\ln\hat\lambda^2 \right]
    + \theta(\hat p^2-\hat\lambda^2)\,
    \frac{4(\hat p^2-\hat\lambda^2)}
         {(\hat p^2-\hat\lambda^2)^2 + z^2\hat\lambda^2} +
    \nonumber\\
   &&{}+ \frac{\theta(\hat p^2-\hat\lambda^2)}{\hat p^2}
    \left[ \left(1 - \frac{\hat\lambda^2}{\hat p^2} \right)
    \left(3 + \frac{\hat\lambda^2}{\hat p^2} \right)
    + 4\ln\left( \frac{\hat\lambda^2}{\hat p^2} 
    + \frac{\hat p^2}{z^2} \right) \right] \,.
\end{eqnarray}
In deriving this expression we have kept the regulator in all terms
that diverge stronger than a logarithm in the limit where $\hat p^2\to 
0$. The terms proportional to $\delta(\hat p^2)$ come from virtual 
corrections to the leading-order diagram, whereas the remaining terms 
arise from cut diagrams with the emission of a real gluon. The limit 
$\hat\lambda\to 0$ can be taken either if $\hat p^2>0$ by virtue of 
some experimental cut, or if the decay distribution is integrated 
over some range in $\hat p^2$. Studying the integrals of the function 
$f_{\rm IR}$ with arbitrary regular test functions $F(\hat p^2)$, we 
find that in the sense of distributions one can replace 
$f_{\rm IR}(\hat\lambda^2,z,\hat p^2)$ by
\begin{equation}\label{fIR}
   f_{\rm IR}(0,z,\hat p^2)=
   \delta(\hat p^2) \left( 4\ln^2\!z - 4\ln z
   + \frac{\pi^2}{3} - \frac52 \right)+ 4\left(
   \frac{\ln\hat p^2}{\hat p^2} \right)_{\!*}
   - (8\ln z-7)\left( \frac{1}{\hat p^2} \right)_{\!*} \,, \qquad
\end{equation}
where the $*$ distributions are defined as
\begin{eqnarray}
    \left( \frac{1}{\hat p^2} \right)_{\!*}
    &=& \lim_{\epsilon\to 0}
     \left[ \frac{\theta(\hat p^2-\epsilon)}{\hat p^2}
     + \delta(\hat p^2) \ln\epsilon \right] \,, \nonumber\\
    \left( \frac{\ln\hat p^2}{\hat p^2} \right)_{\!*}
    &=& \lim_{\epsilon\to 0} \left[ 
     \frac{\theta(\hat p^2-\epsilon)}{\hat p^2}\ln\hat p^2
     + \frac12\,\delta(\hat p^2) \ln^2\!\epsilon \right] \,. 
\end{eqnarray}
This definition is such that 
\begin{equation}
   \int_0^{\hat m^2}\!\mbox{d}\hat p^2\,F(\hat p^2)
   \left( \frac{1}{\hat p^2} \right)_{\!*}
   = F(0)\ln\hat m^2
   + \int_0^{\hat m^2}\!\mbox{d}\hat p^2\,
   \frac{F(\hat p^2)-F(0)}{\hat p^2} \,,
\end{equation}
and similarly for the second distribution, so that the $*$ can be 
omitted if the distributions are multiplied by a test function of 
$O(\hat p^2)$. 

Using the result (\ref{fIR}), the $O(\alpha_s)$ corrections to $W_1$ 
can be rewritten as
\begin{eqnarray}\label{W1new}
   \frac{m_b^2}{2}\,W_1^{(1)} &=& - \delta(\hat p^2) \left[
    8\ln^2\!z - 10\ln z + \frac{2\ln z}{1-z} + 4 L_2(1-z)
    + \frac{4\pi^2}{3} + 5 \right] - \nonumber\\
   &&\mbox{}- 4\left( \frac{\ln\hat p^2}{\hat p^2} \right)_{\!*}
    + (8\ln z-7)\left( \frac{1}{\hat p^2} \right)_{\!*}
    + \frac{4}{\hat p^2} \left[ \frac{1}{t} \ln\frac{1+t}{1-t}
    + \ln\frac{\hat p^2}{z^2} \right] + \nonumber\\
   &&\mbox{}+ 1 - \frac{(8-z)(2-z)}{z^2 t^2} + \left[
    \frac{2-z}{2z} + \frac{(8-z)(2-z)}{2z^2 t^2} \right]
    \frac{1}{t} \ln\frac{1+t}{1-t} \,. \qquad
\end{eqnarray}
The expressions (\ref{Wi}) and (\ref{W1new}) are the basis for all
results presented in this paper.

\section{Double differential distributions}
\label{sec:double}

The exact results for the invariant functions $W_i$ given above allow
the calculation of arbitrary $B\to X_u\,l\,\bar\nu_l$ decay
distributions to next-to-leading order in $\alpha_s$. In particular,
experimental cuts on the variables $E_l$, $E_H$ and $s_H$ can be
implemented in a straightforward way if the distributions are obtained
from a numerical integration of~(\ref{rate}). In this section, we
derive analytic results for the double differential distributions in
the variables $(z,\hat p^2)$ and $(z,x)$ obtained after one
integration of the fully differential decay distribution.  These
results are the basis for, e.g., the study of the hadronic energy
spectrum with a cut on the charged-lepton energy (or vice versa), or
the hadronic invariant mass distribution. The latter will be discussed
in section~\ref{sec:hadmass}.

\subsection{Distribution in the variables $(z,\hat p^2)$}

Inserting the results for the invariant functions $W_i$ into the
general relation~(\ref{double}), we obtain a remarkably simple
expression for the double differential decay rate. Defining
\begin{equation}\label{doub}
   \frac{1}{\Gamma_0}\,
   \frac{\mbox{d}^2\Gamma}{\mbox{d}z\,\mbox{d}\hat p^2}
   = 2 z^2 (3-2z) \left[ \delta(\hat p^2) 
   + \frac{C_F\alpha_s}{4\pi}\,E_1(z,\hat p^2) \right]
   + \frac{C_F\alpha_s}{4\pi}\,E_2(z,\hat p^2) \,,
\end{equation}
where here and below we omit writing $O(\alpha_s^2)$ for the 
neglected higher-order contributions, we find
\begin{eqnarray}
   E_1(z,\hat p^2) &=& - \delta(\hat p^2) \left[
    8\ln^2\!z - 10\ln z + \frac{2\ln z}{1-z} + 4 L_2(1-z) + 5
    + \frac{4\pi^2}{3} \right] - \nonumber\\
   &&{}-4\left( \frac{\ln\hat p^2}{\hat p^2} \right)_{\!*}
    + (8\ln z-7) \left( \frac{1}{\hat p^2} \right)_{\!*} 
    + \frac{1}{\hat p^2} \left[ 8\ln\frac{1+t}{2} + 7(1-t) \right]
    \,, \nonumber\\
   E_2(z,\hat p^2) &=& \frac{4z^3\ln z}{1-z}\,\delta(\hat p^2)
    - 4 \left[ 2z(3-4z) - 3(1-2z)\hat p^2 - 2\hat p^4 \right]
    \ln\frac{1+t}{1-t} + \nonumber\\
   &&{}+4zt\,(10-15z+8\hat p^2) \,. 
\end{eqnarray}
The kinematic range for the variables $z$ and $\hat p^2$ is given 
in (\ref{ps}).

\subsection{Distribution in the variables $(x,z)$}
\label{sec:xz}

Another useful distribution is obtained by integrating the triple
differential decay rate over the variable $\hat p^2$ in the range
specified in~(\ref{ps3}). This leaves $x$ and $z$ as kinematic 
variables, which allows us to compute arbitrary distributions in the 
charged-lepton, neutrino or hadronic energies. The result for this 
distribution takes a different form for the two cases $z<1$ and 
$z>1$. For the first case, we find
\begin{equation}
   \left. \frac{1}{\Gamma_0}\,
   \frac{\mbox{d}^2\Gamma}{\mbox{d}x\,\mbox{d}z}\right|_{z<1}
   = 12(2-x-z)(x+z-1) \left[ 1 + \frac{C_F\alpha_s}{4\pi}\,
   \frac{F_<(x,z)}{2-x-z} \right] \,,
\end{equation}
where $1-z\le x\le 1$, and
\begin{eqnarray}\label{F1}
   F_<(x,z) &=& -2(2-x-z) \left[ \ln^2\!\left( \frac{x+z-1}{1-x}
    \right) + 2 L_2(1-z) + \frac{2\pi^2}{3} \right] - \nonumber\\
   &&{}- 2(5-2x-2z)\ln(1-x) - \frac{f_1}{15}
    \ln\frac{x+z-1}{1-x} + \nonumber\\
   &&{}+ \frac{z f_2}{15(x+z-1)} \ln\frac{z}{1-x}
    - \frac{f_3}{30} \,.
\end{eqnarray}
For the case $z>1$, we have instead
\begin{equation}
   \left. \frac{1}{\Gamma_0}\,
   \frac{\mbox{d}^2\Gamma}{\mbox{d}x\,\mbox{d}z} \right|_{z>1}
   = \frac{C_F\alpha_s}{4\pi}\,F_>(x,z) \,,
\end{equation}
where $0\le x\le 2-z$ and
\begin{eqnarray}\label{F2}
   F_>(x,z) &=& -24(2-x-z)(x+z-1) \left[
    \ln^2\!\left( \frac{x+z-1}{1-x} \right) - \ln^2(z-1) \right] -
    \nonumber\\
   &&{}- \frac{4 f_1}{5}\,(x+z-1) \ln(x+z-1) 
    - \frac{4 f_4}{5} \ln\frac{z-1}{1-x} + \nonumber\\
   &&{}+ \frac{4 f_5}{5}\,(2-x-z)^3 \ln(1-x) 
    + \frac{2x f_6}{5}\,(2-x-z) \,.
\end{eqnarray}
In~(\ref{F1}) and~(\ref{F2}), the coefficients $f_i$ are polynomials 
in $x$ and $z$ given by
\begin{eqnarray}
   f_1 &=& \left(71-44x+26x^2-9x^3+x^4\right) - 
	\left(44-52x+27x^2-4x^3\right)z + 
    \nonumber\\
   &&{}+ \left(26-27x+6x^2\right)z^2 - \left(9-4x\right)z^3 
	+ z^4 \,, \nonumber\\
   f_2 &=& 60x - 15\left(2-3x+5x^2\right)z 
	+ 5\left(7-8x+2x^2\right)z^2 - 5\left(2-x\right)z^3
    + z^4 \,, \nonumber\\
   f_3 &=&  15\left(2+13x-7x^2+3x^3-x^4\right) + 
	\left(37-126x-29x^2-32x^3\right)z -
    \nonumber\\
   &&{}- (1-x)(17-7x)z^2 + 2(1-x)z^3 \,, \nonumber\\
   f_4 &=& \left(71-115x+40x^2\right) - 5\left(23-4x\right)z 
	+ 5\left(14-9x+15x^2\right)z^2 - 
    \nonumber\\
   &&{}- 5\left(7-8x+2x^2\right)z^3 + 5\left(2-x\right)z^4 
	- z^5 \,, \nonumber\\
   f_5 &=& \left(1+4x-x^2\right) + 2\left(2-x\right)z - z^2 \,, \nonumber\\
   f_6 &=& 15\left(10-6x+2x^2-x^3\right) 
	- 2\left(32+23x+16x^2\right)z + 7\left(2-x\right)z^2
    - 2z^3 \,.\ 
\end{eqnarray}
In the limit $x\to 1$, we obtain the simple expression
\begin{equation}
   \left. \frac{1}{\Gamma_0}\,
   \frac{\mbox{d}^2\Gamma}{\mbox{d}x\,\mbox{d}z} \right|_{x\to 1}
   = 12z(1-z) \left[ 1 - \frac{C_F\alpha_s}{2\pi}\,F_1(z,x) 
   \right] \,,
\end{equation}
where
\begin{eqnarray}
   F_1(x,z) &=& \ln^2(1-x) - 2\ln z\ln(1-x) + \frac72 \ln(1-x) + 
    \nonumber\\
   &&{}+ \ln^2\!z - \frac32 \ln z + \frac{\ln z}{1-z}
    + 2 L_2(1-z) + \frac{2\pi^2}{3} + \frac 52 + O(1-x) \,.
\end{eqnarray}
This result, which was previously derived by Akhoury and 
Rothstein~\cite{AkRo}, is needed for the next-to-leading order resummation of 
Sudakov logarithms to all orders of perturbation theory~\cite{Korch,MNnew}.

\section{Single differential spectra}

While the results presented in the previous section are new, there
already exist several calculations of the $O(\alpha_s)$ corrections 
to single differential spectra in $B\to X_u\,l\,\bar\nu_l$ decays. 
Here we derive the distributions in the variables $x$, $z$ and 
$\hat p^2$. This allows comparison of our results with existing 
calculations.

\subsection{Charged-lepton energy spectrum}

Integrating the double differential decay rate derived in 
section~\ref{sec:xz} over $z$ yields the spectrum in the variable $x$, 
which measures the energy of the charged lepton in the $B$-meson 
rest frame. We obtain
\begin{equation}\label{lepten}
   \frac{1}{\Gamma_0}\,\frac{\mbox{d}\Gamma}{\mbox{d}x}
   = 2x^2(3-2x) \left[ 1 - \frac{C_F\alpha_s}{2\pi}\,G(x) \right] \,,
\end{equation}
where $0\le x\le 1$, and
\vspace*{-1.5ex}
\begin{eqnarray}
   G(x) &=& \ln^2(1-x) + 2 L_2(x) + \frac{2\pi^2}{3} 
    + \frac{82-153x+86x^2}{12x(3-2x)} + \nonumber\\
   &&{}+ \frac{41-36x+42x^2-16x^3}{6x^2(3-2x)} \ln(1-x) \,.
\end{eqnarray}
This agrees with the well-known result obtained first by Je\.zabek and 
K\"uhn~\cite{JeKu}. 

{\renewcommand\belowcaptionskip{-1em}\renewcommand\abovecaptionskip{-.3ex}
\EPSFIGURE{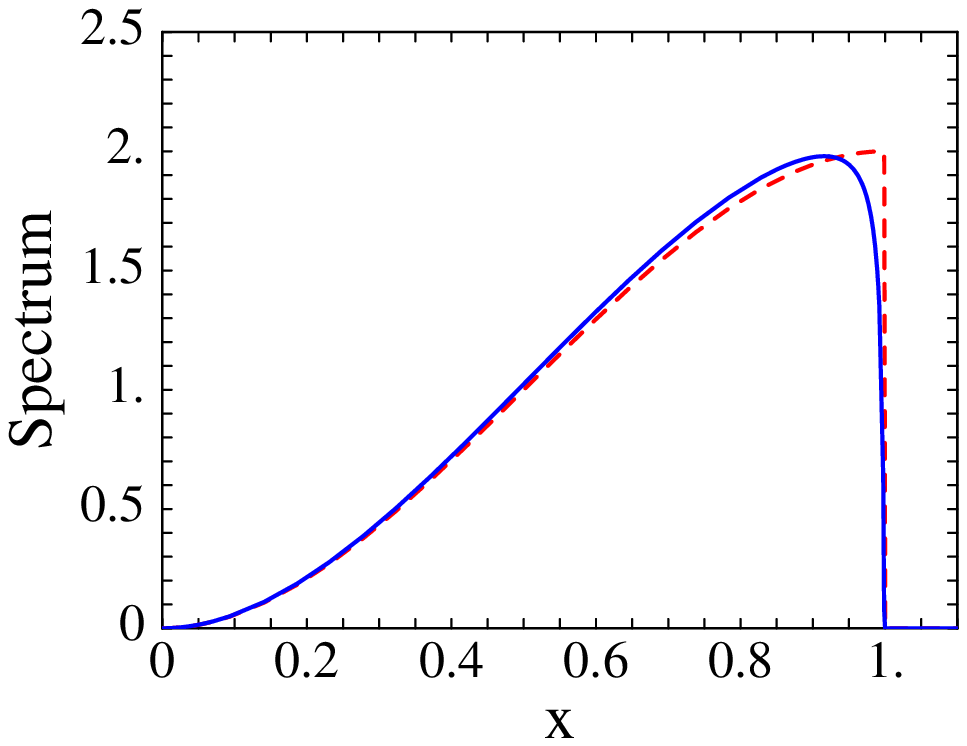,width=17.7em, height=13em}{\label{fig:2}Lepton
energy spectrum $\mbox{d}\Gamma/\mbox{d}x$ in units of $\Gamma$, at
tree level (dashed) and including $O(\alpha_s)$ corrections (solid).}}

The function $G(x)$ is regular for $x\!\to\!0$. On the other hand for
$x\!\to\!1$, i.e.\ close to the boundary of phase space,
there are Sudakov logarithms reflecting the incomplete cancellation of
infrared divergences due to the limited phase space available for real
gluon emission:
\begin{eqnarray}
   G(x) &=& \ln^2(1-x) + \frac{31}{6}\,\ln(1-x) + \nonumber\\[.1ex]
   &&{}+ \pi^2 + \frac54 + O(1-x) \,.
\end{eqnarray}
These endpoint singularities are inte\-gra\-ble, and the total decay 
rate is given by
\begin{equation}\label{gamtot}
   \Gamma = \Gamma_0 \left[ 1 - \frac{C_F\alpha_s}{2\pi} \left(
    \pi^2 - \frac{25}{4} \right) \right] \,.
\end{equation}
In figure~\ref{fig:2}, we show the result for the charged-lepton energy 
spectrum obtained at leading and next-to-leading order in perturbation 
theory, using $\alpha_s=0.22$. Here and below we normalize the 
distributions to the total decay rate $\Gamma$, so that the spectra 
shown have unit area. At tree level we use $\Gamma=\Gamma_0$, whereas 
at next-to-leading order we take the result given in~(\ref{gamtot}). It 
is evident from the figure that the perturbative corrections affect
the spectral shape close to the endpoint only.

\newpage

\subsection{Hadronic energy spectrum}

\EPSFIGURE{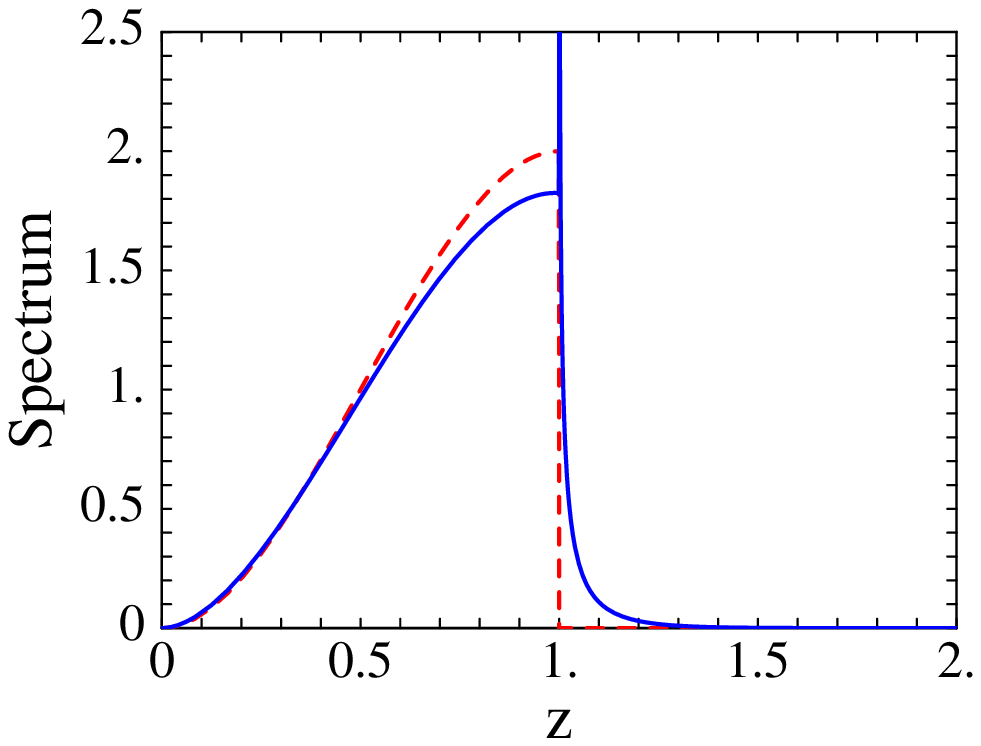,width=7.3cm} {\label{fig:3}Hadronic energy
spectrum $\mbox{d}\Gamma/\mbox{d}z$ in units of $\Gamma$, at tree
level (dashed) and including $O(\alpha_s)$ corrections (solid).}

\noindent Because of the relation
\begin{eqnarray}
   v\cdot p &=& E_H - \bar\Lambda \nonumber\\
   &=& M_B - (E_l + E_{\bar\nu_l}) - \bar\Lambda \,,~~
\end{eqnarray}
both the hadronic energy spectrum and the distribution of the total
energy of the lepton pair can be derived from the distribution for the 
scaling variable $z$, which is obtained by integrating the double 
differential distribution in~(\ref{doub}) over $\hat p^2$.
Because of 
the pha\-se-space con\-straint\linebreak  shown in~(\ref{ps}), the resulting 
expressions are dif\-ferent for the two cases $z<1$ and $z>1$. In the 
second case $\hat p^2=0$ is not allowed by kinematics, and thus only 
diagrams with real gluon emission contribute. We find
\begin{eqnarray}\label{hadren}
   \frac{1}{\Gamma_0}\,\frac{\mbox{d}\Gamma}{\mbox{d}z}
   &=& 2z^2(3-2z) \left[ 1 - \frac{C_F\alpha_s}{2\pi}\,H_<(z)
    \right] \,, \qquad 0\le z\le 1 \,, \nonumber\\
   \frac{1}{\Gamma_0}\,\frac{\mbox{d}\Gamma}{\mbox{d}z}
   &=& \frac{C_F\alpha_s}{2\pi}\,H_>(z) \,, \hspace{4.1cm}
    1\le z\le 2 \,,
\end{eqnarray}
with
\begin{eqnarray}
   H_<(z) &=& 2 L_2(1-z) + \pi^2 + \frac{9-4z}{3-2z}\ln z
    - \frac{4860-3720z+585z^2-42z^3+4z^4}{360(3-2z)} \,, \nonumber\\
   H_>(z) &=& 2z^2(3-2z) \bigg[ \ln^2(z-1) -2\ln^2\!z
    - 4 L_2\!\left( \frac{1}{z} \right) + \frac{\pi^2}{3} \bigg] +
    \nonumber\\
   &&{}+ \frac{(2-z)(1248+3798z-2946z^2+517z^3-34z^4+4z^5)}{180}
    + \nonumber\\
   &&{}+ \frac{5+12z+12z^2-8z^3}{3}\ln(z-1) \,.
\end{eqnarray}
Our results for the functions $H_<(z)$ and $H_>(z)$ agree with the
findings of Czarnecki, Je\.zabek and K\"uhn~\cite{CJK}. Whereas the 
function $H_<(z)$ is regular for $z\to 1$, $H_>(z)$ exhibits a 
logarithmic divergence at this point, because the singularities from 
soft gluon emission are not compensated by virtual gluon corrections. 
We find
\begin{equation}
   H_>(z) = 2\ln^2(z-1) + 7\ln(z-1) - \frac{2\pi^2}{3} 
   + \frac{2587}{180} + O(z-1) \,.
\end{equation}
The result for the hadronic energy spectrum at leading and 
next-to-leading order in perturbation theory is shown in 
figure~\ref{fig:3}. The double-logarithmic singularity at the point 
$z=1$ located inside the allowed kinematic region provides an example 
of a ``Sudakov shoulder''~\cite{Cata}.

\subsection{Parton mass spectrum}

Because the parton invariant mass $\sqrt{p^2}$ is not an observable 
quantity, the spectrum in this variable is not of direct 
phenomenological relevance. However, in the region $p^2=O(m_b^2)$
it follows from~(\ref{rela}) that $p^2\approx s_H$ up to 
corrections of order $\bar\Lambda/M_B$, and thus the spectrum 
$\mbox{d}\Gamma/\mbox{d}p^2$ is a reasonable approximation to the 
had\-ronic invariant mass distribution. Also, this spectrum can be 
used to compute moments of the hadronic mass distribution, since for 
instance $\langle s_H\rangle=m_b^2\,\langle\hat p^2\rangle + 
\bar\Lambda m_b\langle z\rangle + \bar\Lambda^2$. Finally, as we will 
discuss in section~\ref{sec:ope}, the parton mass spectrum allows 
us to perform a quantitative study of the behaviour of perturbative 
QCD in the region of time-like momenta.

Integrating the double differential decay rate in~(\ref{double}) over
$z$ in the range $2\sqrt{\hat p^2}\le z\le 1+\hat p^2$, we obtain
\begin{equation}\label{res1}
   \frac{1}{\Gamma_0}\,\frac{\mbox{d}\Gamma}{\mbox{d}\hat p^2}
   = \delta(\hat p^2) \left[ 1 - \frac{C_F\alpha_s}{2\pi} \left(
   \pi^2 + \frac{187}{72} \right) \right] 
   + \frac{C_F\alpha_s}{2\pi}\,I(\hat p^2) \,,
\end{equation}
where $0\le\hat p^2\le 1$, and 
\begin{eqnarray}\label{res2}
   I(\hat p^2) &=& -2 \left( \frac{\ln\hat p^2}{\hat p^2}
    \right)_{\!*} - \frac{31}{6} \left( \frac{1}{\hat p^2}
    \right)_{\!*} - \left( \frac{10}{3} - 6\hat p^2
    + \frac43\,\hat p^6 \right) \ln\hat p^2 + \nonumber\\
   &&{}+ \frac{67}{9} + 3\hat p^2 - \frac{25}{3}\,\hat p^4
    + \frac{55}{18}\,\hat p^6 \,.
\end{eqnarray}
The regular terms in this result, i.e.~the terms that remain for
$\hat p^2>0$, have been computed previously by Falk, Luke and 
Savage~\cite{FLS}, and we agree with their result.

\section{Hadronic invariant mass distribution}
\label{sec:hadmass}

Imposing a kinematic cut $s_H<M_D^2$ on the inclusive semileptonic 
decay rate of $B$ mesons is an efficient way to separate the 
Cabibbo-suppressed signal from $b\to u$ transitions from the 
background of $b\to c$ decays~\cite{Barg}. Using the relation
between the parton variables $p^2$ and $v\cdot p$ and the hadronic
invariant mass displayed in~(\ref{rela}), and denoting $\hat s_H
=s_H/m_b^2$ and $\varepsilon=\bar\Lambda/m_b$, we find that
\begin{equation}
   \frac{\mbox{d}\Gamma}{\mbox{d}\hat s_H}
   = \int_{z_0}^{z_1}\!\mbox{d}z\,\left. 
   \frac{\mbox{d}^2\Gamma}{\mbox{d}z\,\mbox{d}\hat p^2}
   \right|_{\displaystyle
            \hat p^2=\hat s_H-\varepsilon z-\varepsilon^2} \,,
   \qquad
   \varepsilon^2\le\hat s_H\le(1+\varepsilon)^2 \,,
\end{equation}
where
\begin{equation}
   z_0 = 2(\sqrt{\hat s_H}-\varepsilon) \,, \qquad
   z_1 = \hbox{min}\!\left(
   \frac{\hat s_H-\varepsilon^2}{\varepsilon},\,
   1-\varepsilon+\frac{\hat s_H}{1+\varepsilon} \right) \,.
\end{equation}
Because of the form of $z_1$, one must distinguish the cases where 
$\hat s_H$ is smaller or larger than $\varepsilon(1+\varepsilon)$. In 
the second case, only diagrams with real gluon emission contribute to 
the spectrum. We define
\begin{eqnarray}\label{dGdsH}
   \frac{1}{\Gamma_0}\,
   \frac{\mbox{d}\Gamma}{\mbox{d}\hat s_H}
   &=& T(\hat s_H,\varepsilon)
    - \frac{C_F\alpha_s}{2\pi}\,J_<(\hat s_H,\varepsilon) \,, \qquad 
    \varepsilon^2\le\hat s_H\le\varepsilon(1+\varepsilon) \,,
    \nonumber\\
   \frac{1}{\Gamma_0}\,
   \frac{\mbox{d}\Gamma}{\mbox{d}\hat s_H}
   &=& \frac{C_F\alpha_s}{2\pi}\,J_>(\hat s_H,\varepsilon) \,,
    \hspace{1.76cm}
    \varepsilon(1+\varepsilon)\le\hat s_H\le(1+\varepsilon)^2 \,.
\end{eqnarray}
The exact results for the functions entering this expression read
\begin{eqnarray}\label{master1}
   \varepsilon\,T(\hat s_H,\varepsilon) &=& 2x^2(3-2x) \,,
    \nonumber\\
   \varepsilon\,J_<(\hat s_H,\varepsilon) &=& 2x^2(3-2x) \left[
    \pi^2 + 2 L_2(1-x) - 2 L_2\!\left(-\frac{x}{\varepsilon}\right)
    \right] + \nonumber\\
   &&{}+ \frac13\,(\varepsilon d_1 - x d_2 - 2x^2 d_3)
    (x+\varepsilon) \ln\!\left(1+\frac{x}{\varepsilon}\right) +
    \nonumber\\
   &&{}+ 2x^2(9-4x) \ln x - \frac{x}{3} \bigg( \varepsilon d_1
    + \frac{x}{2}\,d_4 - \frac{x^2}{3}\,d_5 \bigg) \,, \nonumber\\
   \varepsilon\,J_>(\hat s_H,\varepsilon) &=& 2x^2(3-2x) \times
    \nonumber\\
   &&{}\times \left[ \frac{\pi^2}{3} + \ln^2\!y - 2\ln^2\!x
    + 4 L_2\!\left(-\frac{1}{\varepsilon}\right)
    - 4 L_2\!\left(\frac{1}{x}\right)
    - 2 L_2\!\left(-\frac{x}{\varepsilon}\right) \!\right] +
    \nonumber\\
   &&{}+ \frac13\,(\varepsilon d_1 - x d_2 - 2x^2 d_3)
    (x+\varepsilon) \ln\!\left(
    \frac{y+\varepsilon}{1+\varepsilon}\right) + \nonumber\\
   &&{}+ \bigg( 7 + 4y - \frac{y^2}{\varepsilon^2}\,d_6
    + \frac{2y^3}{3\varepsilon^3}\,d_7 \bigg) \ln y + \nonumber\\
   &&{}+ \frac{1-y}{18(1+\varepsilon)}
    (d_8 + x d_9 - 2x^2 d_{10}) \,,
\end{eqnarray}
where
\begin{equation}
   x = \frac{\hat s_H-\varepsilon^2}{\varepsilon} \,, \qquad
   y = \frac{\hat s_H-\varepsilon(1+\varepsilon)}{1+\varepsilon} \,,
\end{equation}
and $d_i$ are polynomials in $\varepsilon$ given in 
equation~(\ref{di}) of the appendix. 

Of key importance to the determination of $|V_{ub}|$ is the 
question which fraction of all $B\to X_u\,l\,\bar\nu_l$ events
have hadronic invariant mass below the charm threshold. To address
this issue, we compute the integral of the spectrum up to a cutoff 
$\hat m^2=s_H^{\rm max}/m_b^2$ defined as
\begin{equation}\label{partial}
   \Gamma(\hat m^2,\varepsilon) \equiv 
   \int_{\varepsilon^2}^{\hat m^2}\!
   \mbox{d}\hat s_H\,\frac{\mbox{d}\Gamma}{\mbox{d}\hat s_H} \,,
   \qquad \varepsilon^2\le\hat m^2\le(1+\varepsilon)^2 \,,
\end{equation}
and write the result in the form
\begin{eqnarray}
   \Gamma(\hat m^2,\varepsilon)
   &=& \Gamma_0 \left[ t(\hat m^2,\varepsilon)
    - \frac{C_F\alpha_s}{2\pi}\,j_<(\hat m^2,\varepsilon)
    \right] \,, \qquad 
    \varepsilon^2\le\hat m^2\le\varepsilon(1+\varepsilon) \,,
    \nonumber\\
   \Gamma(\hat m^2,\varepsilon)
   &=& \Gamma_0 \left[ 1 - \frac{C_F\alpha_s}{2\pi}\,
    j_>(\hat m^2,\varepsilon) \right] \,, \hspace{0.95cm}
    \varepsilon(1+\varepsilon)\le\hat m^2\le(1+\varepsilon)^2 \,. 
\end{eqnarray}
Introducing the variables
\begin{equation}
   \mu = \frac{\hat m^2-\varepsilon^2}{\varepsilon} \,, \qquad
   \varrho = \frac{\hat m^2-\varepsilon(1+\varepsilon)}
                  {1+\varepsilon} \,,
\end{equation} 
we obtain
\begin{eqnarray}\label{master2}
   t(\hat m^2,\varepsilon) &=& \mu^3(2-\mu) \,, \nonumber\\
   j_<(\hat m^2,\varepsilon) &=& \mu^3(2-\mu) \left[ \pi^2
    + 2 L_2(1-\mu) - 2 L_2\!\left( -\frac{\mu}{\varepsilon} \right)
    \right] + \nonumber\\
   &&{}+ \frac{\pi^2}{3} - 2 L_2(1-\mu) + \left( 2\mu
    + \mu^2 + \frac{20}{3}\,\mu^3 - \frac52\,\mu^4 \right) \ln\mu
    + \nonumber\\
   &&{}+ \frac16 \left( \varepsilon^2 e_1
    + \mu\,\varepsilon e_2 - \mu^2 e_3 + \mu^3 e_4 \right)
    (\mu+\varepsilon) \ln\!\left(1+\frac{\mu}{\varepsilon}\right)
    - \nonumber\\
   &&{}- \frac{\mu}{6}\,e_5 - \frac{\mu^2}{12}\,e_6
    - \frac{\mu^3}{18}\,e_7 + \frac{\mu^4}{72}\,e_8 \,, \nonumber\\
   j_>(\hat m^2,\varepsilon) &=& -\mu^3 (2-\mu) \!\left[ 
    \frac{\pi^2}{3} + \ln^2\!\varrho - 2\ln^2\!\mu
    + 4 L_2\!\left(-\frac{1}{\varepsilon}\right)
    - 4 L_2\!\bigg(\frac{1}{\mu}\bigg)
    - 2 L_2\!\left(-\frac{\mu}{\varepsilon}\right) \!\right] +
    \nonumber\\
   &&{}+ \pi^2 + \ln^2\!\varrho 
    - \frac{1}{6\varepsilon(1+\varepsilon)}
    \left( e_9 + \mu e_{10} + \mu^2\varepsilon e_{11}
    - \mu^3 e_{12} \right) \varrho\ln\varrho - \nonumber\\
   &&{}- \frac16 \left( \varepsilon^2 e_1
    + \mu\,\varepsilon e_2 - \mu^2 e_3 + \mu^3 e_4 \right)
    (\mu+\varepsilon) \ln\frac{\varrho+\varepsilon}{1+\varepsilon}
    + \nonumber\\
   &&{}+ \frac{1}{(1+\varepsilon)^2} \left(
    \frac{1}{72}\,e_{13} - \frac{\mu}{18}\,e_{14}
    - \frac{\mu^2}{12}\,e_{15} + \frac{\mu^3}{18}\,e_{16}
    - \frac{\mu^4\varepsilon}{72}\,e_{17} \right) \,,
\end{eqnarray}
with coefficients $e_i$ given in equation~(\ref{ei}) of the appendix.

\FIGURE[t]{\epsfig{file=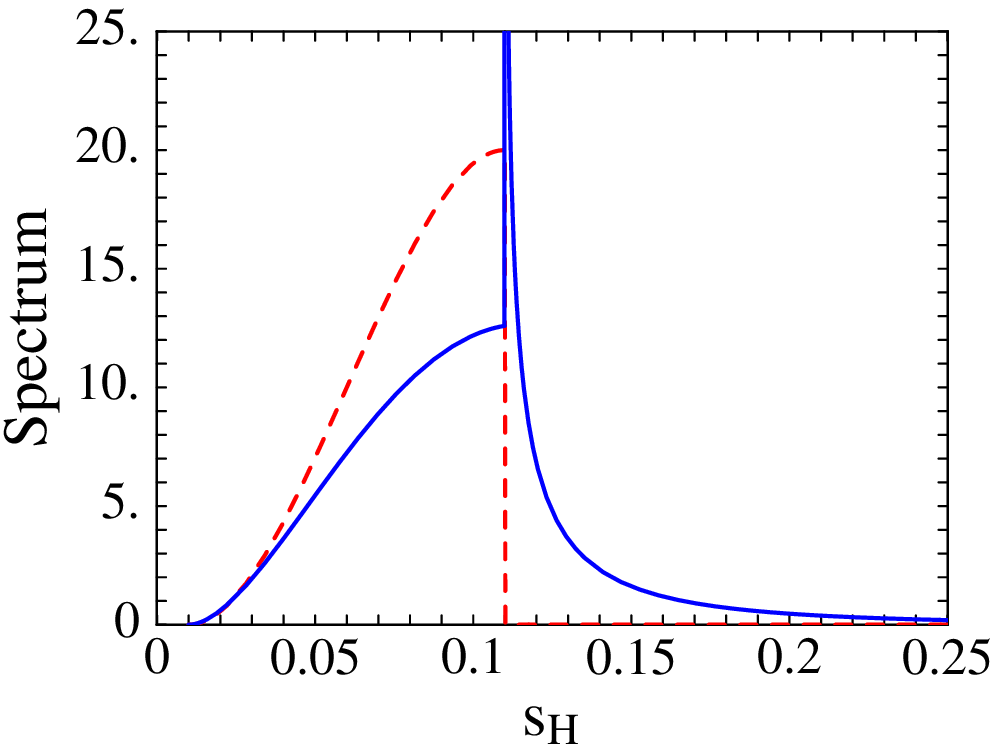,width=17.5em}~
\epsfig{file=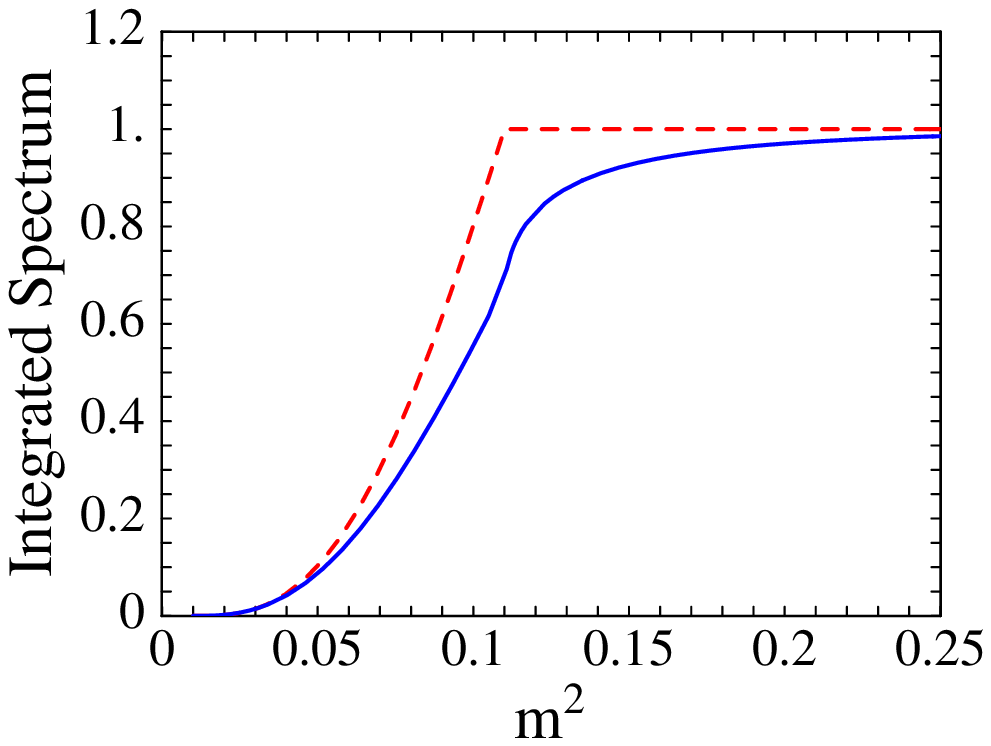,width=17.5em}%
\caption{\label{fig:mass}Left: Invariant hadronic mass spectrum $\mbox{d}\Gamma/\mbox{d}\hat 
s_H$ in units of $\Gamma$, at tree level (dashed) and including 
$O(\alpha_s)$ corrections (solid). Right: Partially integrated 
spectrum $\Gamma(\hat m^2,\varepsilon)$ in units of $\Gamma$.}}

In figure~\ref{fig:mass}, we show the results for the hadronic
invariant mass spectrum (left-hand plot) and for the integral of this 
spectrum up to a cutoff $\hat m^2$ (right-hand plot) for 
$\varepsilon=0.1$, corresponding to $m_b\approx 4.8$\,GeV. Only the 
lower portion of the kinematic range for the variables $\hat s_H$ and 
$\hat m^2$ is displayed. In contrast with the energy spectra 
considered in the previous section, radiative corrections have an 
important impact on the shape of the hadronic invariant mass spectrum 
and lead to a significant redistribution from lower to higher masses. 
Nevertheless, it is apparent from the right-hand plot that imposing a 
cut $s_H<M_D^2$, corresponding to $\hat m^2\approx 0.15$, would leave 
most of the $B\to X_u\,l\,\bar\nu_l$ events unaffected but at the 
same time remove all $B\to X_c\,l\,\bar\nu_l$ 
events~\cite{Barg,Dike,FLW}. Because such a cut falls close to the sharp 
edge of the perturbative hadronic mass spectrum, however, a careful 
treatment of nonperturbative corrections is necessary before any 
conclusions can be drawn. This will be discussed in the following 
section.

\section{Implementation of Fermi motion}
\label{sec:fermi}

The perturbative results presented above are only one ingredient to a
consistent theoretical description of inclusive decay spectra. In
kinematic regions close to phase-space boundaries these spectra are
infrared sensitive and receive large nonperturbative corrections.
Because the corresponding effects can be associated with the motion of
the $b$ quark inside the $B$ meson, they are commonly referred to as
``Fermi motion''. These effects are always important when the 
perturbative prediction for an inclusive decay distribution exhibits a 
rapid variation on a scale that is parametrically smaller than $m_b$. 
Usually, in such a case one encounters large perturbative
corrections from Sudakov logarithms. For the single differential
spectra considered in this paper, this happens in the endpoint
region $1-x=O(\Lambda/m_b)$ of the charged-lepton energy spectrum
shown in figure~\ref{fig:2}, the central region $1-z=O(\Lambda/m_b)$
of the hadronic energy spectrum shown in figure~\ref{fig:3}, and the
low-mass region $\hat s_H=O(\Lambda/m_b)$ of the hadronic invariant 
mass spectrum shown in figure~\ref{fig:mass}. Note that in the first
two cases these are small fractions of the kinematic regions; however,
in the case of the hadronic mass distribution essentially all of the 
spectrum is concentrated in the region $\hat s_H=O(\Lambda/m_b)$, 
where nonperturbative effects are important.

Fermi motion effects are included in the heavy-quark expansion by 
resumming an infinite set of leading-twist corrections into a shape 
function $F(k_+)$, which governs the light-cone momentum distribution 
of the heavy quark inside the $B$ meson~\cite{me,Fermi}. The physical 
decay distributions are obtained from a convolution of parton model 
spectra with this function. In the process, phase-space boundaries 
defined by parton kinematics are transformed into the proper physical 
boundaries determined by hadron kinematics. The shape function is a 
universal characteristic of the $B$ meson governing inclusive decay 
spectra in processes with massless partons in the final state, such 
as $B\to X_u\,l\,\bar\nu_l$ and $B\to X_s\gamma$. The convolution of 
parton spectra with this function is such that in the perturbative 
formulae for the decay distributions the $b$-quark mass $m_b$ is 
replaced by the momentum dependent mass $m_b+k_+$, and similarly 
the parameter $\bar\Lambda=M_B-m_b$ is replaced by 
$\bar\Lambda-k_+$~\cite{me}. 
Here $k_+$ can take values between $-m_b$ and $\bar\Lambda$, 
with a distribution centered around $k_+=0$ and with a characteristic 
width of $O(\Lambda)$. Introducing the new variable 
$q_+=\bar\Lambda-k_+$, it follows that, e.g., the scaling variables 
$x$ and $z$ are replaced by the new variables
\begin{equation}
   x_q = \frac{2E_l}{M_B-q_+} \,,\qquad
   z_q = \frac{2(E_H-q_+)}{M_B-q_+} \,,
\end{equation}
and the physical spectra for the charged-lepton energy and for the 
total hadronic energy are, respectively, given by
\begin{equation}
   \frac{\mbox{d}\Gamma}{\mbox{d}E_l} = 2\!\int_0^{M_B-2E_l}
   \!\!\mbox{d}q_+\,\frac{F(\bar\Lambda-q_+)}{M_B-q_+}\,
   \frac{\mbox{d}\Gamma}{\mbox{d}x}(x_q) \,,\qquad
   0\le E_l\le \frac{M_B}{2} \,,
\end{equation}
and
\begin{equation}\label{EHsp}
   \frac{\mbox{d}\Gamma}{\mbox{d}E_H} = 2\int_0^{E_H}\!
   \mbox{d}q_+\,\frac{F(\bar\Lambda-q_+)}{M_B-q_+}\,
   \frac{\mbox{d}\Gamma}{\mbox{d}z}(z_q) \,,\qquad
   0\le E_H\le M_B \,.
\end{equation}
The perturbative spectra $\mbox{d}\Gamma/\mbox{d}x$ and
$\mbox{d}\Gamma/\mbox{d}z$ have been given in~(\ref{lepten}) 
and~(\ref{hadren}). The upper limits of the $q_+$ integration follow
from the allowed kinematic ranges for the variables $x_q$ and
$z_q$. Similarly, for the hadronic invariant mass spectrum 
we define
\begin{equation}
   \hat s_{H,q}^2 = \frac{s_H}{(M_B-q_+)^2} \,,\qquad
   \varepsilon_q = \frac{q_+}{M_B-q_+} \,,
\end{equation}
and obtain
\begin{equation}\label{sHsp}
   \frac{\mbox{d}\Gamma}{\mbox{d}s_H} = \int_0^{\sqrt{s_H}}
   \!\mbox{d}q_+\,\frac{F(\bar\Lambda-q_+)}{(M_B-q_+)^2}\,
   \frac{\mbox{d}\Gamma}{\mbox{d}\hat s_H}
    (\hat s_{H,q}^2,\varepsilon_q) \,,\qquad
   0\le s_H\le M_B^2 \,,
\end{equation}
with $\mbox{d}\Gamma/\mbox{d}\hat s_H$ as given in~(\ref{dGdsH}).
Here the upper limit of the $q_+$ integration is enforced by 
the requirement that $\hat s_{H,q}^2\ge\varepsilon_q^2$. 
From~(\ref{sHsp}), it follows that the integral over the hadronic 
invariant mass spectrum from 0 to some cutoff $s_H^{\rm max}$ takes 
the form
\begin{equation}\label{magic}
   \Gamma(s_H^{\rm max}) \equiv
   \int_0^{s_H^{\rm max}}\!\mbox{d}s_H\,
   \frac{\mbox{d}\Gamma}{\mbox{d}s_H} 
   = \int_0^{\sqrt{s_H^{\rm max}}}\!\mbox{d}q_+\,
   F(\bar\Lambda-q_+)\,\Gamma(\hat m_q^2,\varepsilon_q) \,,
\end{equation}
where $\hat m_q^2=s_H^{\rm max}/(M_B-q_+)^2$, and the quantity 
$\Gamma(\hat m^2,\varepsilon)$ has been defined in~(\ref{partial}).

After the implementation of Fermi motion the kinematic variables take 
values in the entire phase space determined by hadron kinematics, 
rather than the parton phase space appropriate for perturbative 
calculations. For instance, the maximum lepton energy attainable is 
$E_l^{\rm max}=M_B/2$ rather than $m_b/2$. In the above formulae the 
only reference to the nonperturbative parameter $\bar\Lambda$ resides 
in the shape function. All other mass parameters refer to physical 
hadron masses or energies. We stress that for the numerical evaluation 
of the convolution integrals it is necessary to have explicit analytic 
results for the perturbative spectra. For the important case of the 
hadronic invariant mass distribution, such results have not been 
presented so far in the literature.\footnote{In \protect\cite{FLW} one 
can find a figure showing numerical results for the function 
$\Gamma(\hat m^2,\varepsilon)$ for some particular choices of 
parameters. These results do not permit the computation of the 
physical quantity $\Gamma(s_H^{\rm max})$.}

Several functional forms for the shape function have been suggested 
in the literature. They are subject to constraints on the moments of 
this function, $A_n=\langle k_+^n\rangle$, which are related to the 
forward matrix elements of local operators on the light cone~\cite{me}. 
The first three moments satisfy $A_0=1$, $A_1=0$ and
$A_2=\frac13\mu_\pi^2$, where $\mu_\pi^2$ is the average momentum
squared of the $b$ quark inside the $B$ meson~\cite{FaNe}. For our 
purposes, it is sufficient to adopt the simple form~\cite{Alex}
\begin{equation}
   F(k_+) = N\,(1-x)^a e^{(1+a)x} \,;\qquad
   x = \frac{k_+}{\bar\Lambda} \le 1 \,,
\end{equation}
which is such that $A_1=0$ by construction (neglecting exponentially 
small terms in $m_b/\Lambda$), whereas the condition $A_0=1$ fixes 
the normalization $N$. The parameter $a$ can be related to the second 
moment, yielding $A_2=\frac13\mu_\pi^2=\bar\Lambda^2/(1+a)$. Thus, 
the $b$-quark mass (or $\bar\Lambda$) and the quantity $a$ (or 
$\mu_\pi^2$) are the two parameters of the function. A typical choice 
of values is $m_b=4.8$\,GeV and $a=1.29$, corresponding to 
$\bar\Lambda\approx 0.48$\,GeV and $\mu_\pi^2\approx 0.3$\,GeV$^2$. 
Below, we keep $a$ fixed and consider the three choices $m_b=4.65$, 
4.8 and 4.95\,GeV. The spread of the results provides a realistic 
estimate of the theoretical uncertainty associated with the treatment 
of Fermi motion. This uncertainty could be removed if the shape 
function were extracted, e.g., from a precise measurement of the 
photon energy spectrum in $B\to X_s\gamma$ decays~\cite{Alex}.
 
\FIGURE[t]{\epsfig{file=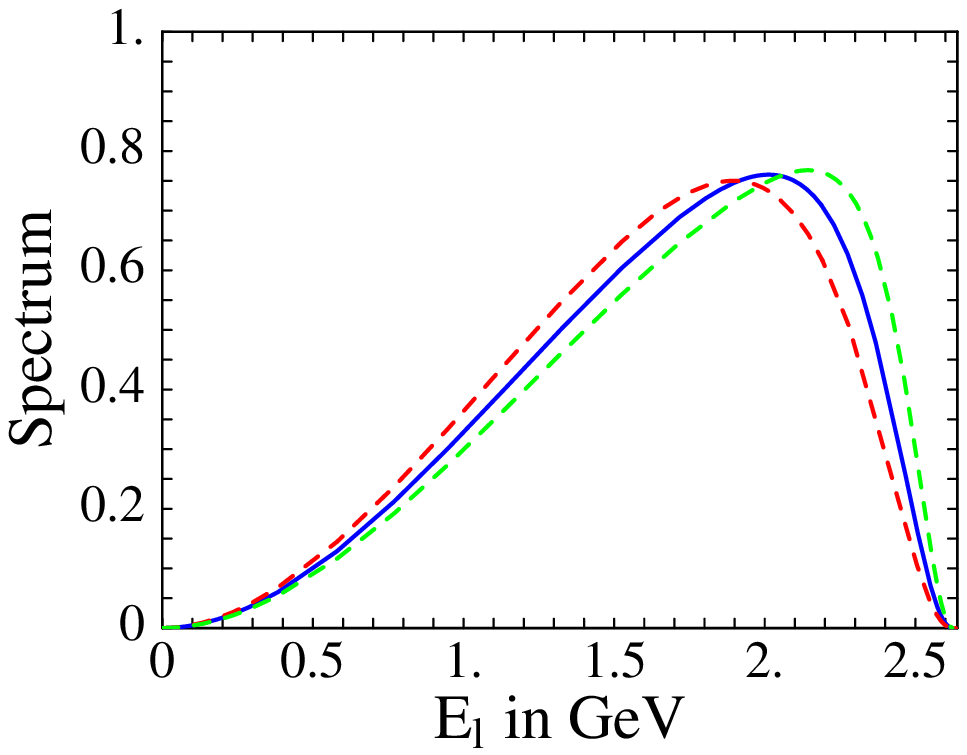,width=7.2cm}
\epsfig{file=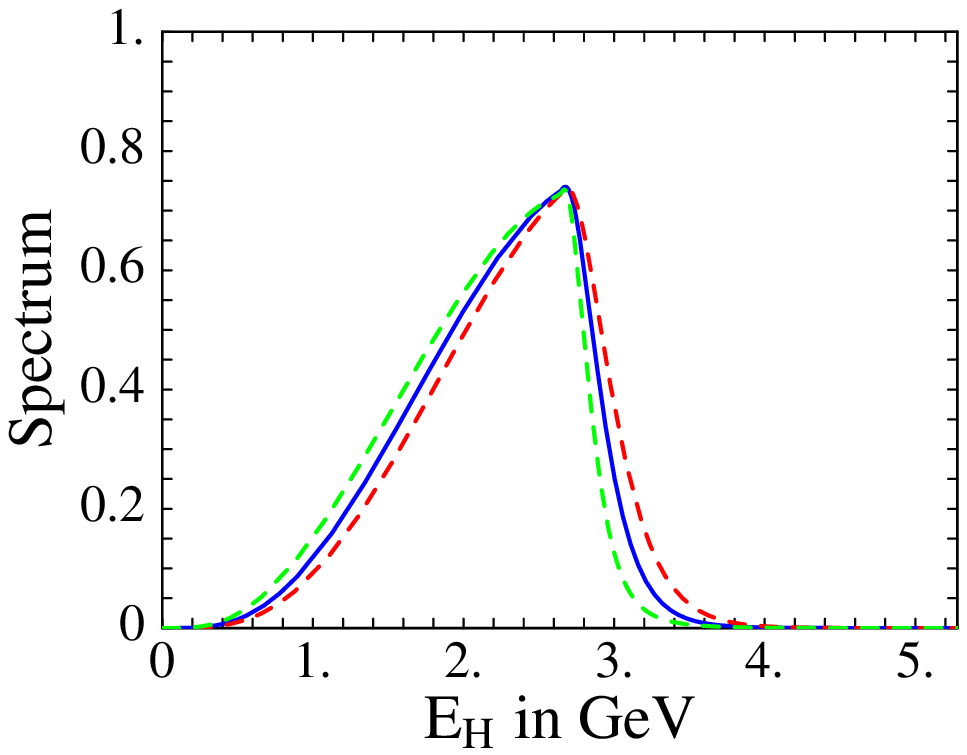,width=7.2cm}%
\caption{\label{fig:spectra}Left: Charged-lepton energy spectrum $\mbox{d}\Gamma/\mbox{d}E_l$ (in 
units of $\Gamma\times\mbox{GeV}^{-1}$) versus $E_l$ for 
$m_b=4.8$\,GeV (solid), 4.65\,GeV (red dashed), and 4.95\,GeV (green 
dashed). Right: Same for the hadronic energy spectrum 
$\mbox{d}\Gamma/\mbox{d}E_H$.}}

Figure~\ref{fig:spectra} shows the results for the charged-lepton 
energy spectrum and for the hadronic energy spectrum, including
Fermi motion effects. The three curves in each plot correspond to 
different values of the $b$-quark mass. Comparing the shape of the 
spectra with the perturbative results shown in figures~\ref{fig:2} 
and \ref{fig:3} indicates that for the charged-lepton energy spectrum 
nonperturbative effects are important in the region $E_l>1.8$\,GeV,
whereas they affect the hadronic energy spectrum in the range 
$2.6\,\mbox{GeV}<E_H<3.6$\,GeV. Generally, Fermi motion effects smooth 
out any sharp structures in the perturbative spectra. From the results 
for the hadronic energy spectrum it follows that the fraction of 
events with $E_H<M_D$ is about $(30\pm 5)\%$, with a moderate 
uncertainty from the dependence on $m_b$. Thus, imposing a cut on the 
energy of the hadronic final state in $B\to X\,l\,\bar\nu_l$ decays 
provides a reasonably efficient way of separating the rare $b\to u$ 
transitions from the much larger background of $B$ decays into 
charmed particles~\cite{Bouz,Greub}.

\FIGURE[t]{\epsfig{file=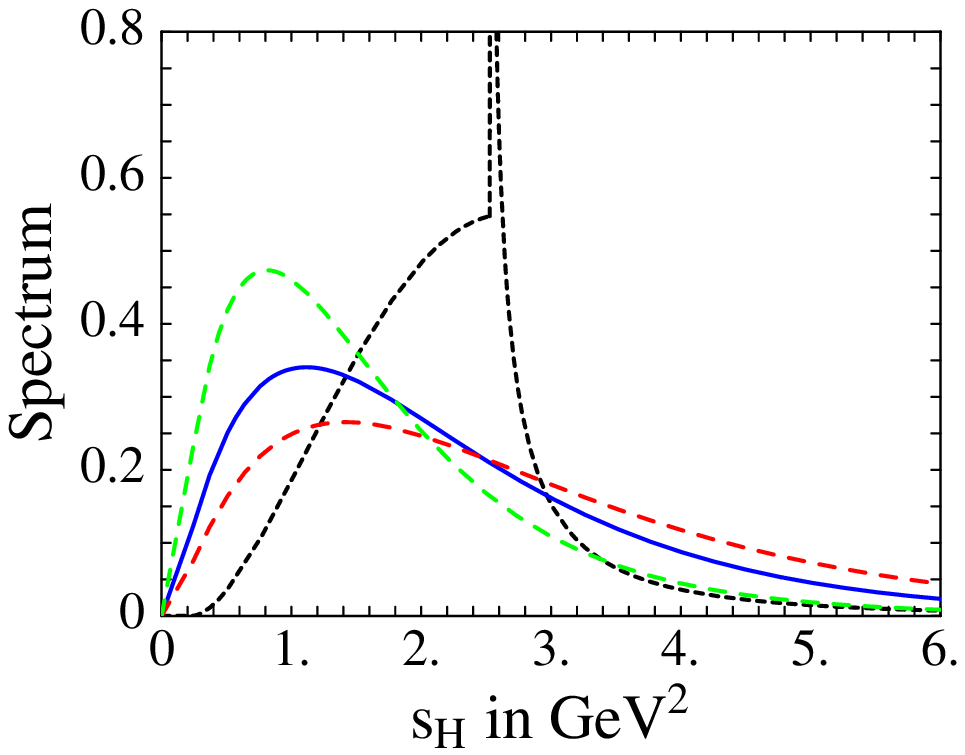,width=7.2cm}
\epsfig{file=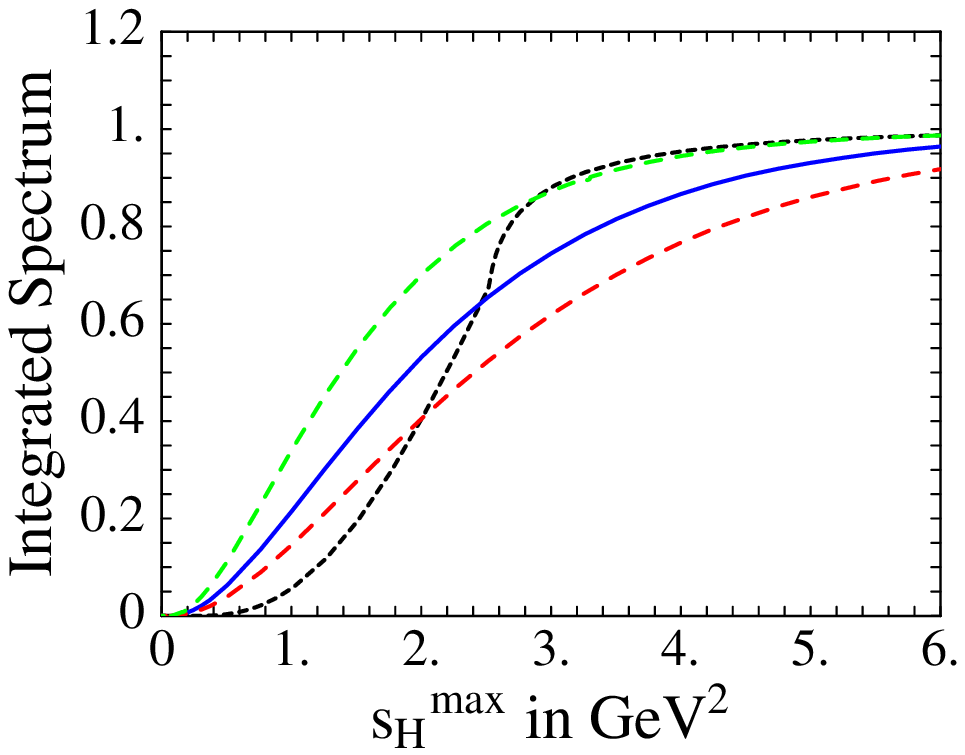,width=7.2cm}%
\caption{\label{fig:integ}Left: Hadronic invariant mass spectrum
$\mbox{d}\Gamma/\mbox{d}s_H$ (in units of $\Gamma\times
\mbox{GeV}^{-2}$) versus $s_H$. The meaning of the curves is the same
as in figure~\protect\ref{fig:spectra}. In addition, the short-dashed
line shows the spectrum for $m_b=4.8$\,GeV without Fermi motion
effects included.  Right: Fraction of $B\to X_u\,l\,\bar\nu_l$ events
with hadronic invariant mass below a cutoff $s_H^{\rm max}$.}}

In Figure~\ref{fig:integ}, we show the results for the hadronic 
invariant mass spectrum and for the fraction of $B\to X_u\,l\,
\bar\nu_l$ events with hadronic mass below a cutoff $s_H^{\rm max}$. 
In each plot, the short-dashed line shows for comparison the 
perturbative spectrum obtained with $m_b=4.8$\,GeV and ignoring the 
effects of Fermi motion. The difference between this curve and the 
solid one is only due to nonperturbative effects. Clearly, these 
effects have a very important impact on the shape of the spectrum in 
the entire low-mass region relevant to experiment, and even in the 
region close to or above the charm threshold. Another important 
observation is that Fermi motion effects remove completely the sharp 
structure at $s_H\sim\bar\Lambda\,M_B$ (corresponding to $\hat s_H
\sim\varepsilon(1+\varepsilon)$) in the perturbative hadronic mass 
spectrum and replace it by a broad bump at a significantly lower 
value of $s_H$. In other words, the value of the unphysical quantity 
$\bar\Lambda$ is of no direct relevance to the hadronic invariant 
mass spectrum.

From the right-hand plot in figure~\ref{fig:integ}, we deduce that 
the fraction of events with hadronic invariant mass below the charm 
threshold ($s_H\le 3.49$\,GeV$^2$) is about $(80\pm 10)\%$. Therefore,
applying a cut $s_H<M_D^2$ would be a most efficient discriminator 
between semileptonc $b\to u$ and $b\to c$ transitions, which would 
allow for a largely model-independent determination of $|V_{ub}|$. If 
for experimental reasons the cutoff on the hadronic mass is lowered to 
$(1.5\,\mbox{GeV})^2$, the fraction of contained events drops to about 
$(60\pm 15)\%$, which is still significant. Even in this more 
pessimistic scenario, $|V_{ub}|$ could be extracted with a theoretical 
uncertainty of about 15\%.

\section{Perturbative QCD in the time-like region}
\label{sec:ope}

As a last application, we use our results for the perturbative
corrections to the inclusive $B\to X_u\,l\,\bar\nu_l$ decay
distributions to investigate the behaviour of perturbative QCD in 
the region of time-like momenta. The heavy-quark expansion used in 
the calculation of inclusive decay rates relies on an application of 
the operator product expansion (OPE) in the Minkowskian region. In 
this region there are physical singularities from on-shell 
intermediate hadron states, which are not reproduced by perturbation 
theory. The hypothesis of quark--hadron duality is the assumption that 
the OPE can still be applied provided there is sufficient averaging 
over hadronic final states, so that the properties of individual hadron 
resonances become unimportant. 

There is at present no known way of quantifying from first principles 
how well this assumption holds in the case of inclusive heavy-quark 
decays. In this section, we investigate the question whether 
perturbation theory itself signals the problem by exhibiting 
singularities when the resonance region is approached. To this end, we 
express the differential decay rate in $(z,\hat p^2)$ as the imaginary 
part of a correlator $T(z,\hat p^2)$,
\begin{equation}
   \frac{1}{\Gamma_0}\,
   \frac{\mbox{d}^2\Gamma}{\mbox{d}z\,\mbox{d}\hat p^2}
   = \frac{1}{\pi}\,\mbox{Im}\,T(z,\hat p^2+i\epsilon) \,,
\end{equation}
where $z$ is taken to be real and inside the interval $0\le z\le 2$. 
Up to irrelevant numerical factors, this correlator is the contraction 
of the tensor $T_{\mu\nu}(v,p)$ defined in~(\ref{Tmunu}) with the 
lepton tensor, integrated over $x$. The imaginary part is nonzero if 
$\hat p^2$ is in the interval given in~(\ref{ps}). It follows that the 
correlator satisfies the dispersion relation
\begin{equation}
   T(z,\hat p^2) = \frac{1}{\pi}
   \int_{z_0}^{z^2/4}\!\mbox{d}\hat s\,
   \frac{\mbox{Im}\,T(z,\hat s+i\epsilon)}{\hat s-\hat p^2} \,,
\end{equation}
where $z_0=\mbox{max}(0,z-1)$. In a similar way, we can express the 
differential spectrum in $\hat p^2$ as the imaginary part of a
correlator $T(\hat p^2)$,
\begin{equation}\label{disp}
   \frac{1}{\Gamma_0}\,
   \frac{\mbox{d}\Gamma}{\mbox{d}\hat p^2}
   = \frac{1}{\pi}\,\mbox{Im}\,T(\hat p^2+i\epsilon) \,, \qquad
   T(\hat p^2) = \frac{1}{\pi}
   \int_0^1\!\mbox{d}\hat s\,
   \frac{\mbox{Im}\,T(\hat s+i\epsilon)}{\hat s-\hat p^2} \,,
\end{equation}
where the imaginary part is nonzero if $0\le\hat p^2\le 1$. With 
this definition, it follows that
\begin{equation}
   T(\hat p^2) = \int_0^2\!\mbox{d}z\,T(z,\hat p^2) \,.
\end{equation}
The correlators $T(z,\hat p^2)$ and $T(\hat p^2)$ as functions
of complex $\hat p^2$ contain information about the behaviour of 
perturbation theory close to the region of physical singularities. 
For the purpose of illustration, we will discuss the structure of 
$T(\hat p^2)$ in detail. An analoguous discussion could be made 
for the correlator $T(z,\hat p^2)$ at any fixed value~of~$z$.

Performing the dispersion integral in~(\ref{disp}) using the 
results for the $O(\alpha_s)$ corrections given in~(\ref{res1})
and~(\ref{res2}), we obtain
\begin{equation}\label{Tp2}
   T(\hat p^2) = - \frac{1}{\hat p^2} \left[ 1 
   - \frac{C_F\alpha_s}{2\pi}\,K(\hat p^2) \right] \,,
\end{equation}
where
\begin{eqnarray}
   K(\hat p^2) &=& - \left( 2 + \frac{10}{3}\,\hat p^2 
    - 6\hat p^4 + \frac43\,\hat p^8 \right)
    L_2\bigg( \frac{1}{\hat p^2} \bigg) + \nonumber\\
   &&{}+ \left( \frac{31}{6} - \frac{41}{18}\,\hat p^2
    - \frac{95}{18}\,\hat p^4 + \frac{55}{18}\,\hat p^6 \right)
    (1-\hat p^2) \ln\!\left(\frac{-\hat p^2}{1-\hat p^2} \right) +
    \nonumber\\
   &&{}+ \pi^2 + \frac{187}{72} - 6\hat p^2
    - \frac{233}{36}\,\hat p^4 + \frac{79}{18}\,\hat p^6 \,.
\end{eqnarray}
The function $K(\hat p^2)$ is analytic in the cut $\hat p^2$ plane
with a discontinuity along the interval $0\le\hat p^2\le 1$, as 
required by the analyticity properties of the correlator following
from~(\ref{disp}). Note that in perturbation theory the only 
singular point is $\hat p^2=0$, where the function has a logarithmic 
singularity. To get a reliable answer for a physical quantity, the 
region close to this singularity must be avoided. 

%\EPSFIGURE{circle.ps,width=5.4cm} 
%{\label{fig:plane}Deformation of the 
%integration contour in the evaluation of the
%quantity $\Gamma(\hat m^2)$.}

Consider, as an example, the inclusive $B\to X_u\,l\,\bar\nu_l$ 
decay rate integrated over an interval in $\hat p^2$ including the 
origin:
\begin{equation}
   \Gamma(\hat m^2) \equiv \int_0^{\hat m^2}\!
   \mbox{d}\hat p^2\,\frac{\mbox{d}\Gamma}{\mbox{d}\hat p^2}
   = \frac{\Gamma_0}{2\pi i} \int_0^{\hat m^2}\!
   \mbox{d}\hat p^2\,\Big[ T(\hat p^2+i\epsilon)
   - T(\hat p^2-i\epsilon) \Big] \,.
\end{equation}
Using Cauchy's relation, the integration contour can be deformed 
into a circle in the complex momentum plane touching the real axis 
at the point $\hat p^2=\hat m^2$. We obtain
\begin{equation}\label{circ}
   \Gamma(\hat m^2) = \Gamma_0 \left[ 1 
   - \frac{C_F\alpha_s}{2\pi} \int_0^{2\pi}
   \frac{\mbox{d}\varphi}{2\pi}\,K(\hat m^2 e^{i\varphi}) \right] \,.
\end{equation}
The situation is illustrated in figure~\ref{fig:plane}. Note that 
for $\hat m^2\ge 1$ the result of the integration is independent
of $\hat m^2$, and it is thus possible to take the radius of the 
circle arbitrarily large, so that the contour is far away from the 
singularities. Therefore, is it generally accepted that the total 
inclusive semileptonic decay rate can be calculated reliably in 
perturbation theory. 

\DOUBLEFIGURE[t]{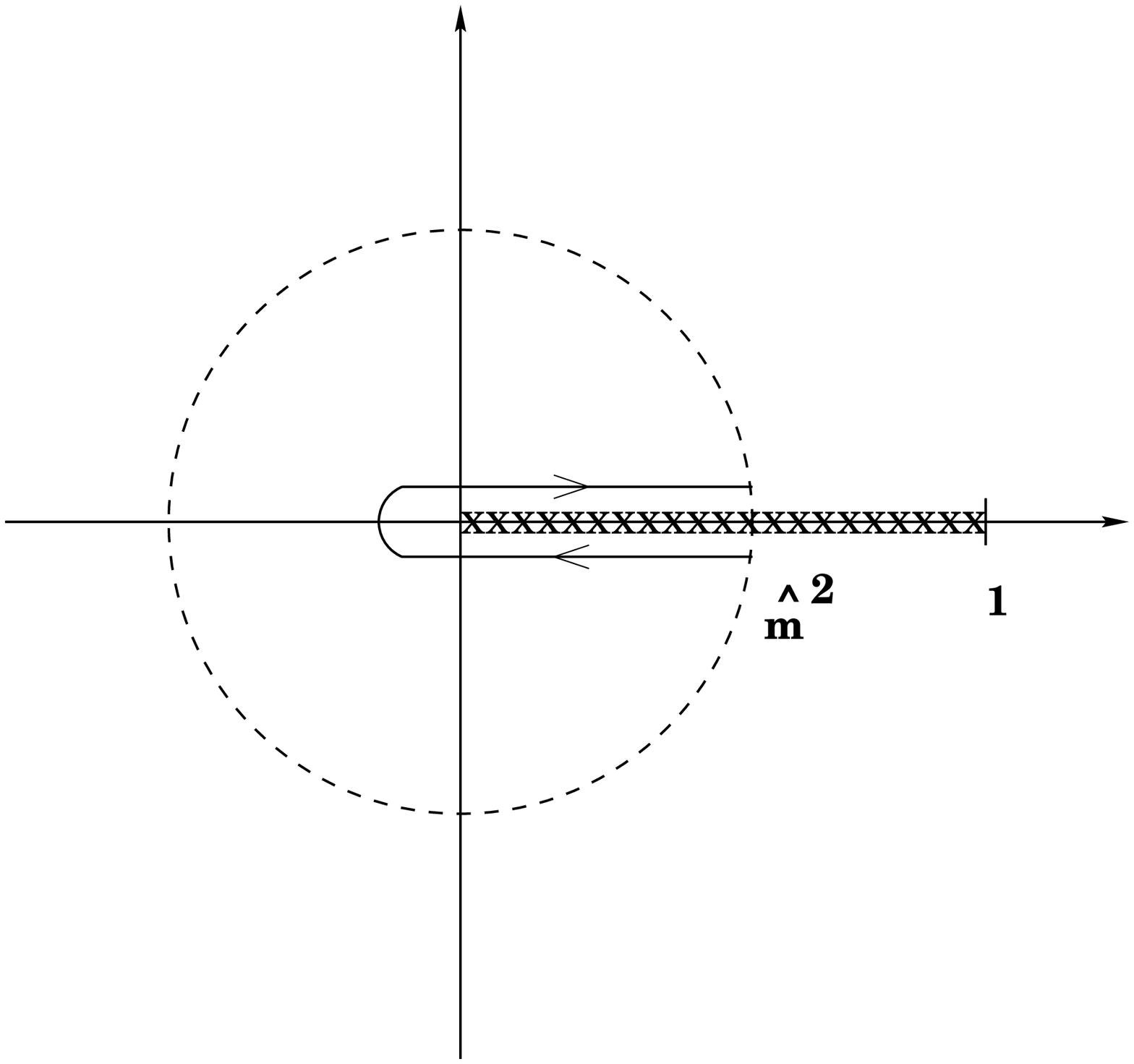,width=7.5cm}{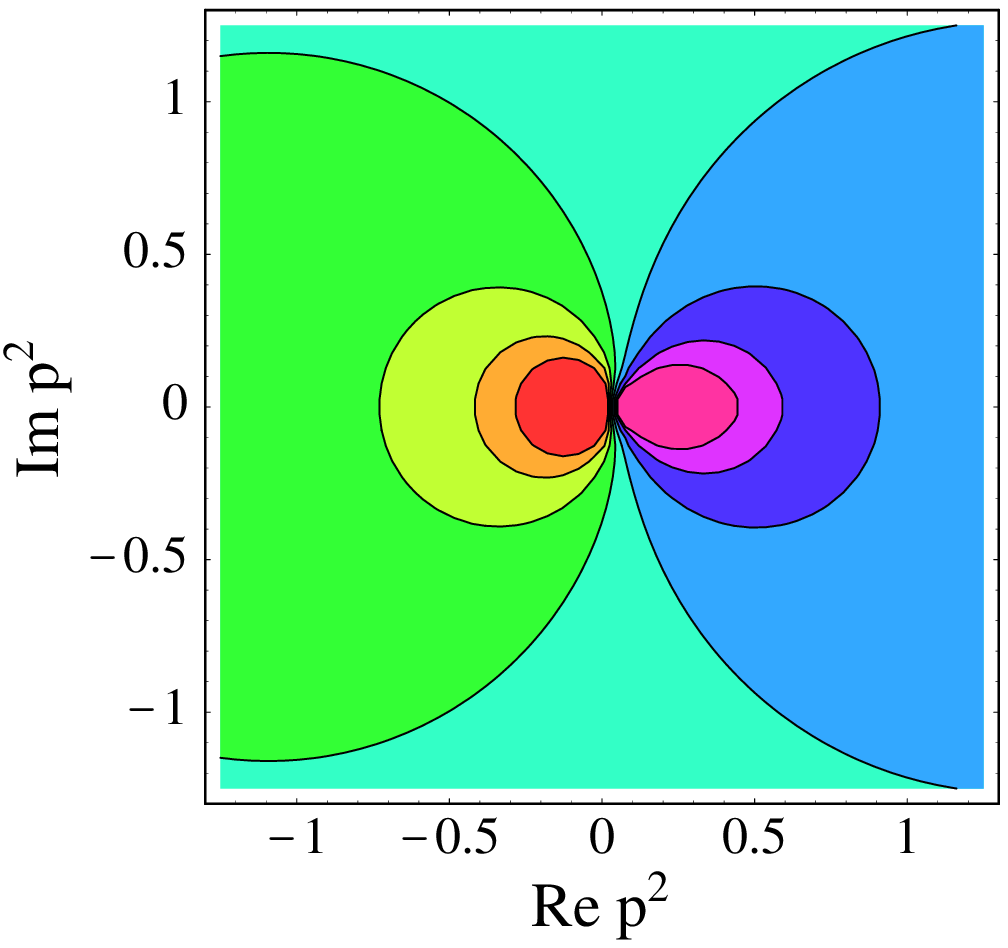,width=6.5cm}
{\label{fig:plane}Deformation of the integration contour in the
evaluation of the quantity $\Gamma(\hat m^2)$.}
{\label{fig:contours}Contours of $\mbox{Re}\,K(\hat p^2)$ in the
com\-plex $\hat p^2$ plane.  The contour lines refer to values between
2.5 (innermost right contour) and 4.5 (innermost left contour) in
units of 0.25.}

If $\hat m^2<1$, on the other hand, the contour probes the region 
of physical resonances at the point where it touches the cut. It is 
natural to ask whe\-ther perturbation theory exhibits singularities in 
the vicinity of the cut, which could signal a breakdown of quark--hadron 
duality in the calculation of partially integrated decay rates such as 
$\Gamma(\hat m^2)$. Remarkably, we find that this is not the case. As 
long as $\hat m^2$ is not too close to the origin, the perturbative 
corrections to the correlator $T(\hat p^2)$ are well behaved everywhere 
in the complex momentum plane, even close to the cut. This is evident 
from figure~\ref{fig:contours}, which shows contours of the real 
part of $K(\hat p^2)$ in the complex $\hat p^2$ plane.\footnote{The 
imaginary part does not contribute to the integral along the circle 
in~(\protect\ref{circ}).} 
We take this as an indication that partially integrated decay rates can 
be calculated using the heavy-quark expansion as long as they are not 
restricted to a range in $\hat p^2$ too close to the origin. 

\section{Conclusions}

%\EPSFIGURE{contours.eps,width=6.5cm} 
%{\label{fig:contours}
%Contours of $\mbox{Re}\,K(\hat p^2)$ in the com\-plex $\hat p^2$ plane.
%The contour lines refer to values between 2.5 (innermost right contour) 
%and 4.5 (innermost left contour) in units of 0.25.}

We have presented analytic results for the next-to-leading order
perturbative corrections to the triple differential 
$B\to X_u\,l\,\bar\nu_l$ decay rate. They provide the basis for
the computation of arbitrary inclusive decay distributions to
$O(\alpha_s)$, including experimental cuts on various kinematic
variables. Our results are sufficiently general to allow treating 
the case where the mass of the charged lepton cannot be neglected. 
As an application, we have presented explicit results for several 
double and single differential distributions, most of which had 
not been derived previously. In particular, we have discussed in 
detail the $O(\alpha_s)$ corrections to the hadronic invariant mass 
distribution, which are an important ingredient in a theoretically 
clean determination of the element $|V_{ub}|$ of the quark mixing 
matrix. 

We have shown how the leading nonperturbative corrections
affecting inclusive decay spectra can be incorporated in a QCD-based
framework by convoluting the perturbative spectra with a $b$-quark
momentum distribution function. This is important for addressing the
question of how experimentally one may separate the $B\to
X_u\,l\,\bar\nu_l$ signal from the large background of semileptonic 
decays into charmed particles. We find that $(30\pm 5)\%$ of all $B\to
X_u\,l\,\bar\nu_l$ events have hadronic energy below the charm
threshold, while $(80\pm 10)\%$ have hadronic invariant mass below
$M_D^2$. If the cutoff on the hadronic mass is lowered to
$(1.5\,\mbox{GeV})^2$, this fraction drops to $(60\pm 15)\%$, which
would still allow for a largely model-independent determination of 
$|V_{ub}|$. 

Finally, we have studied the behaviour of perturbative QCD in the 
complex momentum plane, finding that there is no evidence for large
corrections except for the region close to $p^2=0$. This observation 
can be taken as circumstantial evidence in support of global 
quark--hadron duality, which underlies the heavy-quark expansion
for inclusive decay rates.

\acknowledgments
We would like to thank Pietro Colangelo, Zoltan Ligeti, Aneesh 
Manohar, Giusep\-pe Nardulli, Nello Paver and Helen Quinn for useful 
discussions. The research of~M.N.~was supported by the Department of 
Energy under contract DE--AC03--76SF00515.

\appendix
\section{Coefficients entering the hadronic mass spectrum}

Here we list the coefficients $d_i$ and $e_i$ entering the exact 
theoretical expressions for the hadronic invariant mass spectrum 
in~(\ref{master1}), and for the integrated spectrum in~(\ref{master2}). 
These coefficients are polynomials in $\varepsilon=\bar\Lambda/m_b$
given by
\begin{eqnarray}\label{di}
   d_1 &=& 9 + 10\varepsilon -7\varepsilon^2 - 24\varepsilon^3
    - 40\varepsilon^4 \,, \nonumber\\
   d_2 &=& 9 - 20\varepsilon - 7\varepsilon^2 - 6\varepsilon^3
    +32\varepsilon^4 \,, \nonumber\\
   d_3 &=& 1 + 8\varepsilon - 3\varepsilon^2 + 2\varepsilon^3
    \,, \nonumber\\
   d_4 &=& 153 + 50\varepsilon + 7\varepsilon^2 - 12\varepsilon^3
    - 104\varepsilon^4 \,, \nonumber\\
   d_5 &=& 167 + 34\varepsilon - 39\varepsilon^2 + 40\varepsilon^3
    \,, \nonumber\\
   d_6 &=& 3 + 8\varepsilon + 4\varepsilon^2 \,, \nonumber\\
   d_7 &=& 1 - \varepsilon - 6\varepsilon^2 - 4\varepsilon^3 \,, 
    \nonumber\\
   d_8 &=& 143 + 172\varepsilon + 113\varepsilon^2
    + 42\varepsilon^3 - 146\varepsilon^4 - 384\varepsilon^5
    - 240\varepsilon^6 \,, \nonumber\\
   d_9 &=& 348 + 497\varepsilon + 174\varepsilon^2
    - 7\varepsilon^3 - 388\varepsilon^4 - 312\varepsilon^5 \,,
    \nonumber\\
   d_{10} &=& 138 + 159\varepsilon - 4\varepsilon^2 + 17\varepsilon^3
    + 40\varepsilon^4 \,,
\end{eqnarray}
and
\begin{eqnarray}\label{ei}
   e_1 &=& 4 - 12\varepsilon^2 - 25\varepsilon^3 - 30\varepsilon^4
    \,, \nonumber\\
   e_2 &=& 14 + 20\varepsilon - 2\varepsilon^2 
    - 23\varepsilon^3 - 50\varepsilon^4 \,, \nonumber\\
   e_3 &=& 14 - 10\varepsilon - 2\varepsilon^2
    - 5\varepsilon^3 + 22\varepsilon^4 \,, \nonumber\\
   e_4 &=& 2 - 8\varepsilon + 3\varepsilon^2 - 2\varepsilon^3 \,,
    \nonumber\\
   e_5 &=& 12 + 4\varepsilon^2 - 12\varepsilon^4
    - 25\varepsilon^5 - 30\varepsilon^6 \,, \nonumber\\
   e_6 &=& 6 + 32\varepsilon + 40\varepsilon^2 - 16\varepsilon^3
    - 71\varepsilon^4 - 130\varepsilon^5 \,, \nonumber\\
   e_7 &=& 179 + 60\varepsilon + 9\varepsilon^2 - 7\varepsilon^3 
    - 126\varepsilon^4 \,, \nonumber\\
   e_8 &=& 373 + 92\varepsilon - 87\varepsilon^2
    + 86\varepsilon^3 \,, \nonumber\\
   e_9 &=& 26 + 40\varepsilon + 17\varepsilon^2 \,, \nonumber\\
   e_{10} &=& 18 + 44\varepsilon + 19\varepsilon^2 \,, \nonumber\\
   e_{11} &=& 8 + 13\varepsilon \,, \nonumber\\
   e_{12} &=& 2 + 8\varepsilon + 7\varepsilon^2 \,, \nonumber\\
   e_{13} &=& 13 + 446\varepsilon + 340\varepsilon^2
    - 160\varepsilon^3 - 26\varepsilon^4 + 572\varepsilon^5
    + 1428\varepsilon^6 + 1680\varepsilon^7 + 720\varepsilon^8 \,,
    \nonumber\\
   e_{14} &=& 65 + 338\varepsilon + 394\varepsilon^2
    + 244\varepsilon^3 - 38\varepsilon^4 - 529\varepsilon^5
    - 732\varepsilon^6 - 225\varepsilon^7 + 90\varepsilon^8 \,,
    \nonumber\\
   e_{15} &=& 98 + 292\varepsilon + 265\varepsilon^2
    + 24\varepsilon^3 - 141\varepsilon^4 - 216\varepsilon^5
    + 43\varepsilon^6 + 130\varepsilon^7 \,, \nonumber\\
   e_{16} &=& 78 + 396\varepsilon + 396\varepsilon^2
    + 83\varepsilon^3 + 37\varepsilon^4 - 115\varepsilon^5
    - 126\varepsilon^6 \,, \nonumber\\
   e_{17} &=& 312 + 372\varepsilon + 4\varepsilon^2
    + 37\varepsilon^3 + 86\varepsilon^4 \,.
\end{eqnarray}

\end{document}